\documentclass[11pt,letterpaper]{article}

\usepackage[T1]{fontenc}
\usepackage[cp1251]{inputenc}
\usepackage{textcomp}
\usepackage[centertags]{amsmath}
\usepackage{amsfonts}
\usepackage{amssymb}
\usepackage{graphicx}
\usepackage[pdftex]{hyperref}
\usepackage[numbers,sort&compress]{natbib}
\usepackage{paperinitial}

\paperinitialization{15mm}{15mm}{20mm}{10mm}{2pt}{10pt}

\DeclareMathOperator{\sgn}{sgn}
\DeclareMathOperator{\Tr}{Tr}
\DeclareMathOperator{\re}{Re}

\newcommand{\e}{\varepsilon}

\newcommand{\s}{\sigma}

\newcommand{\al}{\alpha}
\newcommand{\be}{\beta}
\newcommand{\ga}{\gamma}
\newcommand{\Ga}{\Gamma}
\newcommand{\de}{\delta}

\newcommand{\la}{\lambda}

\newcommand{\spx}{\mathbf{x}}
\newcommand{\spy}{\mathbf{y}}
\newcommand{\spp}{\mathbf{p}}

\newcommand{\R}{\mathbb{R}}
\newcommand{\N}{\mathbb{N}}

\begin{document}
\frenchspacing
\allowdisplaybreaks[4]

\title{{\Large \textbf{One-loop omega-potential of charged massive particles in a constant homogeneous magnetic field at high temperatures}}}

\date{}

\author{I.S. Kalinichenko${}^{1)}$\thanks{E-mail: \texttt{theo@sibmail.com}},\; P.O. Kazinski${}^{1),2)}$\thanks{E-mail: \texttt{kpo@phys.tsu.ru}}\\[0.5em]
{\normalsize ${}^{1)}$ Physics Faculty, Tomsk State University, Tomsk 634050, Russia}\\
{\normalsize ${}^{2)}$ Department of Higher Mathematics and Mathematical Physics,}\\
{\normalsize Tomsk Polytechnic University, Tomsk 634050, Russia}}

\maketitle

\begin{abstract}

The explicit expressions for the high-temperature expansions of the one-loop corrections to the omega-potential coming from charged scalar and Dirac particles and, separately, from  antiparticles in a constant homogeneous magnetic field are derived. The explicit expressions for the non-perturbative corrections to the effective action at finite temperature and density are obtained. Thermodynamic properties of a gas of charged scalars in a constant homogeneous magnetic field are analyzed in the one-loop approximation. It turns out that, in this approximation, the system suffers a first-order phase transition from the diamagnetic to the superconducting state at sufficiently high densities. The improvement of the one-loop result by summing the ring diagrams is investigated. This improvement leads to a drastic change in thermodynamic properties of the system. The gas of charged scalars passes to the ferromagnetic state in place of the superconducting one at high densities and sufficiently low temperatures, in the high temperature regime.

\end{abstract}

\section{Introduction}

We revisit the classical problem of thermodynamic behaviour of a gas of charged particles in a strong constant homogeneous magnetic field at the one-loop level \cite{Schafroth,KirzhLind,Linde,Ditt,Kapustp,ShabadDT,ChodEverOw,LoewRoj,ElPerSka93,ElPerSka94,DFGK,PerRoj1,KirTom,ElmSkag95,ELPS}. In the case of scalars, there are many controversies in the literature regarding the properties of such a gas at high temperatures and densities. In \cite{Schafroth}, it was shown using the naive one-loop approximation that a gas of charged scalars passes to the superconducting state at sufficiently high densities. Later, this result was confirmed in many papers both in the non-relativistic and relativistic domains \cite{DFGK,ELPS,PerRoj1,PerRoj2,PerRojVilLel,DelBarSol,StandTom2,Toms1995,DaicFr96,StandTom1}. However, it is astonishing that the order of this phase transition remains unknown. In some papers \cite{PerRoj1,PerRoj2,PerRojVilLel,DelBarSol,DaicFr96}, the authors suggested that this is the ``diffusive'' type of the phase transition (crossover) without the critical temperature. In other papers, there exist claims that, in a $3$-dimensional space, a gas of charged scalars does not condense (in the sense of the existence of phase transition) at any temperature and density provided the magnetic field is not zero \cite{KirTom,StandTom2,Toms1995,StandTom1}, no matter how it is small. In the paper \cite{ELPS}, it was shown that, in any finite \emph{local} magnetic field, the Bose–Einstein condensation of a relativistic boson gas does not happen, but this Bose gas can condense in the non-zero \emph{external} magnetic field. Nevertheless, the critical temperature and the order of this phase transition were not found in \cite{ELPS}. There is another group of papers \cite{ALRV,AHLMRV} where the authors suggested that such a gas can condense if one goes beyond the naive one-loop approximation and takes into account the infrared enhanced contribution of the so-called ring diagrams \cite{KirzhLind,Linde,DolJack,KapustB}.

As for the naive one-loop approximation, the conclusions following from our study in this paper mainly agree with those given in \cite{ELPS}. We, however, establish that, in this approximation, such a gas of bosons behaves at high temperatures and densities as the usual superconductor of the first type. The phase transition from the normal to superconducting state is first-order with the definite critical temperature which we also find. If one considers the relativistic Ginzburg-Landau model in the state where the gauge symmetry is not spontaneously broken and takes into account the contribution of the ring diagrams to the omega-potential then instead of the superconducting state the ferromagnetic phase arises at high temperatures and densities. The phase transition to the ferromagnetic state is first-order, and we derive the formula for the Curie temperature in this model. This behaviour takes place for any positive self-interaction coupling constant, when the perturbation theory makes sense, and for the physical value of the fine structure constant. So, we may infer that the high-temperature superconductivity discussed in \cite{DFGK,ELPS,PerRoj1,PerRoj2,PerRojVilLel,DelBarSol,Toms1995,DaicFr96} is just an artefact of the naive one-loop approximation.

As a byproduct of our investigation, we verify the general formulas for the high-temperature expansion derived in \cite{KalKaz1,KalKaz2}. Furthermore, these general formulas allow us to obtain the high-temperature expansions of the one-loop corrections to the omega-potential from particles and antiparticles separately, i.e., to generalize the results of \cite{DFGK,CanDunne}. Such formulas are necessary, for example, in considering the number of particles created by heating of the system (see, e.g., \cite{KhalilB}), the total charge of the system being maintained constant. We derive such high-temperature expansions for both charged scalars and charged Dirac fermions in a constant homogeneous magnetic field. The explicit formulas for the non-perturbative corrections to the effective action at finite temperature are also obtained. These corrections are non-analytic in the coupling constant and cannot be reproduced by a straightforward summation of Feynman diagrams. Their form resembles the instanton contributions to the effective action (see, for review, \cite{GuilZinJust}).

So, we start in Sect. \ref{Gen_Form_HTE} with a brief reminder of the general formulas for the high-temperature expansion derived in \cite{KalKaz1,KalKaz2}. In Sect. \ref{Heat_Kernel}, we derive the explicit expression for the heat kernel associated with the Klein-Gordon equation in a constant homogeneous electromagnetic field at finite temperature and density. In spite of the fact that the main subject of the paper is related to the homogeneous magnetic field only, the general formulas obtained in Sect. \ref{Heat_Kernel} lay down the basis for further investigations both in the one-loop and higher loop calculations, where off-diagonal elements of the heat kernel are necessary. In Sect. \ref{Const_Hom_Field}, we apply the general formulas and derive the explicit expressions for all the elements of the high-temperature expansion for scalar and Dirac particles. In particular, we obtain there the strong and weak field expansions of the one-loop omega-potential including the non-perturbative corrections. The explicit formulas for the high-temperature expansions are collected in Sect. \ref{High-Temp_Expan}. As known (see, e.g., \cite{KazShip}), the zero temperature contribution to the effective action can be found from the high-temperature expansion. So, we find the zero temperature effective action -- the particular case of the Heisenberg-Euler effective action -- in this section too.

Sect. \ref{Char_Bos_Mag_Field} is devoted to the analysis of thermodynamic properties of a gas of charged bosons in a constant homogeneous magnetic field at high temperatures in the naive one-loop approximation. We describe the isochoric and adiabatic processes in the normal (diamagnetic) phase. As for superconductivity, we establish the main properties of the phase transition from the normal to the superconducting phase. In particular, we find numerically the dependence $H(B)$ and the equilibrium curve of the diamagnetic and superconducting phases. The approximate formulas for the main characteristics of the phase transition are also obtained. In Sect. \ref{Ring_Diag}, we take into account the self-interaction of charged bosons and  photons by means of the ring diagrams and improve the one-loop approximation considered in the previous section. The ring diagrams are taken into account with the aid of the gap equation \cite{DolJack,KirzhLind,Kapustp,Weldon862} on the temperature dependent effective masses. Then it turns out that, instead of the Landau diamagnetism, the particles (photons and scalars) possess effectively the paramagnetic properties. Hence, at high densities, it is energetically favorable for the system to increase the magnetic field rather than to expulse it. Numerical analysis reveals that, at high densities and sufficiently low temperatures (but in the high temperature limit), the system passes to the ferromagnetic state\footnote{Notice that the ferromagnetism of the gas of vector bosons at low temperatures was predicted in \cite{PerRoj1,PerRojVilLel,Khalilp,KhalHo}.}. We describe numerically the dependence $H(B)$, which displays the typical hysteresis loop, and the dependence of the spontaneous magnetization on temperature. The approximate formula for the Curie temperature is also derived. We show numerically that the ferromagnetic state can be reached adiabatically by increasing the temperature provided the entropy per unit charge is not very large.

\section{General formulas for the high-temperature expansion}\label{Gen_Form_HTE}

The one-loop correction to the omega-potential of quantum particles is defined in the standard way
\begin{equation}\label{omega_pot_defn}
    \mp\beta\Omega_{f,b}=\sum_k\ln(1\pm e^{-\be(\omega^{(+)}_k-\mu)})=\pm\be\int_0^\infty d\omega\frac{\Tr\theta(H(\omega))}{e^{\be(\omega-\mu)}\pm1},\qquad \omega^{(+)}_k>0,
\end{equation}
where the upper sign corresponds to fermions $f$, the lower sign is for bosons $b$, and $\omega^{(+)}_k$ is the energy of particles. The contribution from antiparticles has the form \eqref{omega_pot_defn} with the replacements $\mu\rightarrow-\mu$ and $\omega^{(+)}_k\rightarrow\omega^{(-)}_k$, where $\omega^{(-)}_k>0$ is the energy spectrum of antiparticles. The high-temperature expansion of \eqref{omega_pot_defn} in $d$-dimensional space takes the form \cite{KalKaz1,KalKaz2}
\begin{multline}\label{expan_ferm}
    -\Omega_{f}(\mu)=\sum_{k,n=0}(1-2^{2\nu+k+n-d})\Gamma(d+1-2\nu-k)\zeta(d+1-2\nu-k-n)\frac{\zeta_{k}(\nu)(\beta\mu)^n}{n!\beta^{d+1-2\nu-k}}+\\
    +\sum^{\infty}_{l=0}(1-2^{1+l})\frac{(-1)^l \zeta(-l)}{\Gamma(l+1)}\sigma^l_{\nu}(\mu)\beta^l,\quad \nu\rightarrow0;
\end{multline}
\begin{multline}\label{expan_bos}
    -\Omega_{b}(\mu)=\sum_{k,n=0}\Gamma(d+1-2\nu-k)\zeta(d+1-2\nu-k-n)\frac{\zeta_{k}(\nu)(\beta\mu)^n}{n!\beta^{d+1-2\nu-k}}+\\
    +\sum^{\infty}_{l=-1}\frac{(-1)^l \zeta(-l)}{\Gamma(l+1)}\sigma^l_{\nu}(\mu)\beta^l,\quad \nu\rightarrow0;
\end{multline}
up to the terms exponentially suppressed at $\be\rightarrow+0$. The following notation was introduced in \eqref{expan_ferm}, \eqref{expan_bos}. Let us define the function
\begin{equation}\label{zeta_func}
    \zeta_{+}(\nu,\omega)=\int_{C}\frac{d\tau \tau^{\nu-1}}{2\pi i}\Tr e^{-\tau H(\omega)},
\end{equation}
where the contour $C$ goes downwards parallel to the imaginary axis and slightly to the left of it. The operator $H(\omega)$ is the Fourier transform with respect to time of the wave operator. It is the Laplace type operator, and the common sign is chosen such that its spectrum is bounded from above for ``good'' background fields. In the case when the spectral density of $H(\omega)$ does not possess nonintegrable singularities, $\zeta_{+}(\nu,\omega)$ is an entire function of $\nu$ for $\re\nu<1$. For other $\nu\in \mathbb{C}$, the function $\zeta_{+}(\nu,\omega)$ is defined by the analytical continuation. The functions $\sigma^l_{\nu}(\mu)$ are defined as
\begin{equation}\label{sigma_gen}
    \sigma^l_{\nu}(\mu)=\int^{\infty}_{0}d\omega(\omega-\mu)^l \zeta_{+}(\nu,\omega).
\end{equation}
It follows from the derivation of \eqref{expan_ferm}, \eqref{expan_bos} \cite{KalKaz2} that the integration contours in the $\omega$ plane can be rotated a little bit and a proper domain of variable $\nu$ in the complex plane can be chosen in order to provide a convergence of the integrals in \eqref{sigma_gen}. The coefficients $\zeta_{k}(\nu)$ are the coefficients of the asymptotic expansion
\begin{equation}\label{zeta_knu}
    \zeta_{+}(\nu,\omega)=\sum_{k=0}^{\infty}\zeta_{k}(\nu)|\omega|^{d-2\nu}\omega^{-k},\quad\omega\in\R,
\end{equation}
which is obtained when one employs in \eqref{zeta_func} the standard heat kernel expansion of $\Tr e^{-\tau H(\omega)}$ near $\tau=0$ and evaluates the integral over $\tau$ (see for details \cite{KalKaz1,KalKaz2} and below).

As a rule, the coefficients of the heat kernel expansion and, consequently, the coefficients \eqref{zeta_knu} can be easily found. The first six coefficients of the heat kernel expansion for an arbitrary background are given in \cite{Ven} (for their adaptation to the derivation of \eqref{zeta_knu} see \cite{KalKaz1}). The non-trivial problem in deducing the explicit expression for the high-temperature expansions \eqref{expan_ferm}, \eqref{expan_bos} is to find the non-perturbative expression for the diagonal of the heat kernel and to calculate the integrals in \eqref{sigma_gen}. As for the contribution from antiparticles to the one-loop omega-potential, the formulas \eqref{expan_ferm}, \eqref{expan_bos}, \eqref{sigma_gen} are modified in an obvious way \cite{KalKaz2}.

The high-temperature expansion of the one-loop correction to the omega-potential of fermions allows one to obtain the one-loop correction to the effective action both at zero and non-zero chemical potential. The contribution of one bosonic degree of freedom to the non-renormalized one-loop effective action at zero temperature is written as (see, e.g., \cite{KazShip,gmse})
\begin{equation}\label{eff_act_1b}
  \Ga^{(1)}_{1b}/T=-\lim_{\be\rightarrow+0}\partial_\be(\be\Omega_f)_{\mu=0} = \partial_\be\bigg[\be\int_0^\infty d\omega\frac{\Tr\theta(H(\omega))}{e^{\beta\omega}+1}\bigg]_{\be\rightarrow+0},
\end{equation}
where $T$ is a time interval. The non-renormalized omega-potential of Dirac fermions at zero temperature and $\mu\neq0$ with the vacuum contribution takes the form
\begin{equation}\label{eff_act_f}
    \Omega^{(1)}=2\partial_\be\bigg[\be\int_0^\infty d\omega\frac{\Tr\theta(H(\omega+|\mu|))}{e^{\beta(\omega+|\mu|)}+1}\bigg]_{\be\rightarrow+0},\quad\mu\in\R,\;\mu\neq0.
\end{equation}
Recall that the Lagrangian of the effective action $L_{eff}^{(1)}=-\Omega^{(1)}$. In particular, it follows from \eqref{eff_act_1b}, \eqref{eff_act_f} that, in the high-temperature expansion of the total one-loop omega-potential, the vacuum contribution is canceled out by the analogous term in the omega-potential coming from real particles (for fermions in QED see \cite{ElmSkag} and for the general case see \cite{KazShip,KalKaz1,KalKaz2}).

The following stability conditions are assumed in formulas \eqref{omega_pot_defn}, \eqref{eff_act_1b} and \eqref{eff_act_f}:
\begin{enumerate}
  \item The spectrum of $H(\omega)$ does not contain zero eigenvalues at $\omega=0$;
  \item The spectral density $\sgn(\omega)\partial_\omega\Tr \theta(H(\omega))$ is an even non-negative function of $\omega$.
\end{enumerate}
The last condition is satisfied under rather general assumptions about the form of the background fields both for bosons and fermions (see \cite{DeWGAQFT}, Sects. 17, 19).

\section{Heat kernel}\label{Heat_Kernel}

Let us derive the exact expression for the heat kernel entering into \eqref{zeta_func} in the case of a charged massive scalar field on a constant homogeneous electromagnetic background (see also \cite{ShabadDT,ElmSkag95,ElPerSka93,ElPerSka94,Shovkovy,CanDunne}). The heat kernel is an evolution operator
\begin{equation}
    G(\omega,s;\spx,\spy):=\langle \spx|e^{-is(-H(\omega))}|\spy \rangle,
\end{equation}
taken at the imaginary time $s=i\tau$, of some quantum-mechanical system with the ``Hamiltonian'' $-H(\omega)$. In the case at hand
\begin{equation}
    H=-\eta^{\mu\nu}(\partial_{\mu}-iA_{\mu})(\partial_{\nu}-iA_{\nu})-m^2\equiv(p-A)^2-m^2,\quad\eta_{\mu\nu}=diag(1,-1,-1,-1),
\end{equation}
where the charge $e$ is included into the definition of the electromagnetic potential. Therefore,
\begin{equation}\label{h_omega}
    H(\omega)=-(\spp-\mathbf{A})^2-m^2+(\omega-A_{0})^2.
\end{equation}
We expand the electromagnetic field into a series
\begin{equation}
    A_{\mu}(x)=A_{\mu}(0)+\partial_{i}A_{\mu}(0)x^i+\frac{1}{2}\partial_{i}\partial_{j}A_{\mu}(0)x^{i}x^{j}+\dots,
\end{equation}
use the Fock gauge $A_{i}(x)x^{i}=0$, and keep only the linear part in $x$:
\begin{equation}\label{kaliblin}
    A_{i}(x)\approx\frac{1}{2}x^{j}F_{ji},\quad A_{0}(x)\approx A_{0}-E_{i}x^i.
\end{equation}
All the quantities on the right-hand side are taken at the point $x=0$. The potential \eqref{kaliblin} corresponds to the constant electromagnetic field. The constant $A_0(0)$ can be always included into the definition of the chemical potential (see, e.g., \cite{Kapustp}) conjugate to the total electric charge of the system. Henceforth, we put $A_0(0)=0$.

The Hamiltonian \eqref{h_omega} with the field \eqref{kaliblin} is quadratic. Let
\begin{equation}
    H_0(\omega):=-H(\omega)=\Big(p_{i}-\frac{1}{2}x^{j}F_{ji} \Big)^2-2\omega E_{i}x^i-E_{i}E_{j}x^ix^j-\omega^2+m^2.
\end{equation}
The exact expression for the evolution operator generated by the quadratic Hamiltonian is known
\begin{equation}\label{heatker}
    G(\omega,s;\spx,\spy)=\Big[(-2\pi i)^{-d}\det \frac{\partial^2 S}{\partial x^{i}\partial y^{j}}\Big]^{1/2}e^{iS(\omega,s;\spx,\spy)},
\end{equation}
where $S$ is the Hamilton-Jacobi action,
\begin{equation}\label{action}
    S=\int_0^s d\tau(p_i\dot{x}^i-H_0(\omega)),
\end{equation}
evaluated on the trajectory satisfying the boundary conditions $\spx(0)=\spy$, $\spx(s)=\spx$. The immediate derivation of the action $S$ encounters the computational problems related to the necessity to solve the equations of motion
\begin{equation}
    \ddot{x}_{i}-2F_{ij}\dot{x}^j-4E_iE_jx^j-4E_i\omega=0.
\end{equation}
In the case of the constant electromagnetic field of a general configuration, the answer is very cumbersome. Therefore, we find, at first, the action for the four-dimensional problem $S_{4d}$, which can be readily calculated, and relate it to the action \eqref{action}.

Let us introduce
\begin{equation}\label{S_4d}
    S_{4d}(s;x^{\mu}(0),x^{\mu}(s))=\int_{0}^{s}d\tau(p_{\mu}\dot{x}^{\mu}-H_{4d}),
\end{equation}
where
\begin{equation}
    H_{4d}=-(p-A(x))^2+m^2.
\end{equation}
The solution to the equations of motion in the four-dimensional problem with the boundary conditions $x(s)=x$, $x(0)=y$ takes the simple matrix form
\begin{equation}
    x(\tau)=y+\frac{e^{-2F\tau}-1}{e^{-2Fs}-1}(x-y).
\end{equation}
We have for the action
\begin{equation}\label{S_4dpr}
    S_{4d}=x^{\mu}(-\frac14\dot{x}_{\mu}+A_{\mu})|_{0}^{s}-\frac{1}{2}\int_{0}^{s}d\tau(F_{\mu\alpha}+2\partial_{\alpha}A_{\mu})x^{\mu}\dot{x}^{\alpha}-m^2s.
\end{equation}
The action \eqref{S_4dpr} in the gauge $A'_{\mu}=\frac12F_{\nu\mu}x^{\nu}$ is written as
\begin{equation}
    S'_{4d}=-\frac{1}{4}(x-y)F\coth Fs(x-y)-\frac12xFy-m^2s.
\end{equation}
In order to obtain the action \eqref{S_4d} in the gauge \eqref{kaliblin}, we perform the gauge transformation
\begin{equation}
    A_{\mu}=A'_{\mu}-\partial_{\mu}\phi, \qquad S_{4d}=S'_{4d}-\phi|_{0}^s,
\end{equation}
where $\phi=\frac12F_{0i}x^{0}x^{i}$. Then
\begin{equation}
    S_{4d}(s;x^0-y^0,\spx,\spy)=-\frac{1}{4}(x-y)F\coth sF(x-y)-\frac12x^{i}F_{ij}y^{j}-\frac12(x^0-y^0)F_{0i}(x^{i}+y^{i})-m^2s.
\end{equation}

Now we find the action \eqref{action}. Since $A_{\mu}(x)$ does not depend on time in the gauge \eqref{kaliblin}, $p_0$ is an integral of motion of the model \eqref{S_4d}. So, we put $p_0\equiv\omega=const$. It follows from the Hamilton equations for \eqref{S_4d} that
\begin{equation}
    p_0=\omega=\Big(-\frac{F}{2}-\frac{F}{2}\coth Fs \Big)_{0\nu}(x-y)^{\nu}-F_{0\nu}y^{\nu}.
\end{equation}
Whence
\begin{equation}\label{x^0-y^0}
    x^0-y^0=-\Big(\frac{F}{2}\coth Fs\Big)^{-1}_{00}\Big[\omega+\Big(\frac{F}{2}\coth Fs\Big)_{0i}(x^i-y^i)+\frac12F_{0i}(x^i+y^i)\Big].
\end{equation}
In virtue of the fact that on the solutions to the equations of motion
\begin{equation}
    S_{4d}=\int_{0}^{s}d\tau[\omega\dot{x}^0+p_{i}\dot{x}^{i}-H_0(\omega)]=\omega(x^0-y^0)+S,
\end{equation}
we deduce
\begin{equation}
    S(\omega,s;\spx,\spy)=S_{4d}(s;x^0-y^0,\spx,\spy)-\omega(x^0-y^0),
\end{equation}
where, on the right-hand side, $x^0-y^0$ must be replaced by \eqref{x^0-y^0}. Finally, we have
\begin{multline}
    S=(F\coth sF)^{-1}_{00}\Big[\omega+\frac12\Big(F\coth sF\Big)_{0i}(x^i-y^i)+\frac12F_{0i}(x^i+y^i)\Big]^2-\\
    -\frac14(x^i-y^i)(F\coth sF)_{ij}(x^j-y^j)-\frac{1}{2}x^iF_{ij}y^j-m^2s.
\end{multline}
In order to find the Van Vleck determinant, we employ the formula for the determinant of the block matrix
\begin{multline}
    \det\frac{\partial^2S}{\partial x^{i}\partial y^{j}}=\Big(\frac{F}{2}\coth sF-\frac{F}{2}\Big)_{ij}-\Big(\frac{F}{2}\coth sF-\frac{F}{2}\Big)_{i0}\Big(\frac{F}{2}\coth sF-\frac{F}{2}\Big)^{-1}_{00}\Big(\frac{F}{2}\coth sF-\frac{F}{2}\Big)_{0j}=\\
    =\Big(\frac{F}{2}\coth sF-\frac{F}{2}\Big)^{-1}_{00}\det\Big(\frac{F}{2}\coth sF-\frac{F}{2}\Big)_{\mu\nu}
\end{multline}
Taking all together, we obtain the heat kernel \eqref{heatker}
\begin{multline}\label{kernel}
    G(\omega,s;\spx,\spy)=\frac{1}{(-2\pi i)^{3/2}}
    \sqrt{\frac{\det\big(\frac{F}{2}\coth sF-\frac{F}{2}\big)}
    {\big(\frac{F}{2}\coth sF\big)_{00}}}\times\\
    \times e^{i\Big\{(F\coth sF)^{-1}_{00}\big[\omega+\frac12(F\coth sF)_{0i}(x^i-y^i)+\frac12F_{0i}(x^i+y^i)\big]^2-
    \frac14(x^i-y^i)(F\coth sF)_{ij}(x^j-y^j)-\frac{1}{2}x^iF_{ij}y^j-m^2s\Big\}}.
\end{multline}
In particular, on the diagonal (for the Dirac fields see \cite{ElmSkag95,ElmSkag})
\begin{equation}\label{kernel_diag}
    G(\omega,s;\spx,\spx)=\frac{1}{(-2\pi i)^{3/2}}
    \sqrt{\frac{\det\big(\frac{F}{2}\coth sF-\frac{F}{2}\big)}
    {\big(\frac{F}{2}\coth sF\big)_{00}}}
    e^{i\big[(F\coth sF)^{-1}_{00}(\omega+E_ix^i)^2
    -m^2s\big]}.
\end{equation}
The heat kernel in the case $A_0(0)\neq0$ is obtained by substitution $\omega\rightarrow\omega-A_0(0)$ in \eqref{kernel}, \eqref{kernel_diag} .

The expression \eqref{kernel} can be derived in a different way. Let us define the four-dimensional kernel
\begin{multline}
    G_{4d}(s;x,y)=\Big[\frac{1}{(-2\pi i)^4}\det \frac{\partial^2S_{4d}}{\partial x^{\mu}\partial y^{\nu}} \Big]^{1/2} e^{iS_{4d}}=\\
    =\frac{1}{(2\pi)^{2}}\Big[\det\Big(\frac{F}{2}\coth sF-\frac{F}{2}\Big)\Big]^{1/2}e^{-i\Big[\frac{1}{4}(x-y)F\coth sF(x-y)+\frac12x^{i}F_{ij}y^{j}+\frac12(x^0-y^0)E_{i}(x^{i}+y^{i})+m^2s\Big]}.
\end{multline}
The Fourier transform of \eqref{kernel} over $\omega$ gives precisely $G_{4d}$:
\begin{equation}
    \int_{-\infty}^{\infty} \frac{d\omega}{2\pi} e^{i\omega(x^0-y^0)}G(\omega,s;\spx,\spy)=G_{4d}(s;x,y).
\end{equation}
The integral over $\omega$ is Gaussian and is easily evaluated.

In the expression \eqref{kernel}, the following matrix,
\begin{equation}
(F\coth sF)_{\mu\nu}=
\begin{bmatrix}
(F\coth sF)_{00} & (F\coth sF)_{01} & 0 & 0 \\
(F\coth sF)_{10} & (F\coth sF)_{11} & 0 & 0 \\
0 & 0 & (F\coth sF)_{22} & (F\coth sF)_{23} \\
0 & 0 & (F\coth sF)_{32} & (F\coth sF)_{33}, \\
\end{bmatrix},
\end{equation}
and the determinant,
\begin{equation}\label{determ}
    \det\Big(\frac{F}{2}\coth sF-\frac{F}{2}\Big)=\Big(\frac{p_+p_-}{4\sin sp_{-}\sinh sp_{+}}\Big)^2,
\end{equation}
appear. Here
\begin{equation*}
    (F\coth sF)_{00}=\frac{(p_+^2-\mathbf{E}^2)p_-\cot(sp_-)+(p_-^2+\mathbf{E}^2)p_+\coth(sp_+)}{p_+^2+p_-^2},
\end{equation*}
\begin{equation*}
    (F\coth sF)_{11}=-\frac{(p_+^2-\mathbf{E}^2)p_+\coth(sp_+)+(p_-^2+\mathbf{E}^2)p_-\cot(sp_-)}{p_+^2+p_-^2},
\end{equation*}
\begin{equation*}
    (F\coth sF)_{22}=\frac{p_+p_-}{\mathbf{E}^2(p_+^2+p_-^2)}[(p_+^2-\mathbf{E}^2)p_-\coth(sp_+)-(p_-^2+\mathbf{E}^2)p_+\cot(sp_-)],
\end{equation*}
\begin{equation*}
    (F\coth sF)_{33}=\frac{(p_+^2-\mathbf{E}^2)p_-^3\cot(sp_-)-(p_-^2+\mathbf{E}^2)p_+^3\coth(sp_+)}{\mathbf{E}^2(p_+^2+p_-^2)},
\end{equation*}
\begin{equation*}
    (F\coth sF)_{10}=\frac{H_2E_3[p_+\coth(sp_+)-p_-\cot(sp_-)]}{p_+^2+p_-^2},
\end{equation*}
\begin{equation*}
    (F\coth sF)_{23}=H_2H_3\frac{p_-\cot(sp_-)-p_+\coth(sp_+)}{p_+^2+p_-^2},
\end{equation*}
and
\begin{equation}
    p_{\pm}=\big(\sqrt{I_1^2+I_2^2}\pm I_1\big)^{1/2}, \qquad I_1=\frac12(\mathbf{E}^2- \mathbf{B}^2), \quad I_2=(\mathbf{E}\mathbf{B}),
\end{equation}
where $\mathbf{E}$ is the electric field strength and $\mathbf{B}$ is the magnetic induction vector. All these expressions are written in the system of coordinates where $\mathbf{E}=(0,0,E_3)$ and $\mathbf{B}=(0,B_2,B_3)$. The expression for $(F\coth sF)_{00}$ coincides with that found in \cite{ElmSkag,Shovkovy} (see also \cite{ShabadDT}).

All the singularities of the expression \eqref{kernel} in the complex  $s$ plane lie on the imaginary and real axes only. Indeed, the singularities of the heat kernel coincide with   singularities of the determinant \eqref{determ} and the matrix $F\coth sF$, and zeros of $(F\coth sF)_{00}$. The singularities of the determinant and $F\coth sF$ are the poles lying on the real and imaginary axes and corresponding to the solution of the equation $\sin sp_- \sinh sp_{+}=0$. The zeros of $(F\coth sF)_{00}$ are determined by the condition
\begin{equation}\label{singul_eqn}
    \cot(sp_-)\tanh(sp_{+})=a,
\end{equation}
where $a$ is some real number. Substituting $s=x+iy$ and taking the imaginary part of \eqref{singul_eqn}, we obtain that either
\begin{equation}\label{singul_eqn1}
    \frac{\sin 2p_-x}{\sinh 2p_{+}x}=\frac{\sinh 2p_-y}{\sin 2p_{+}y},
\end{equation}
or $x$ or $y$ vanish. However, equality \eqref{singul_eqn1} is only possible for $x=y=0$. Therefore, all the singularities of the expression \eqref{kernel} lie only on the imaginary and real axes in the complex $s$ plane.

\section{Constant homogeneous magnetic field}\label{Const_Hom_Field}
\subsection{Charged scalar field}

Further, we shall investigate thoroughly the case of the constant homogeneous magnetic field $\mathbf{B}$. In this case, the diagonal of the heat kernel becomes
\begin{equation}\label{HK_bos}
    G(\omega,i\tau;\spx,\spx)=\frac{e^{3\pi i/2}}{(4 \pi)^{3/2}\tau^{1/2}}\frac{B}{\sinh\tau B}e^{-\tau(\omega^2-m^2)},
\end{equation}
where $\mathbf{B}=(0,0,B)$. Without loss of generality, we assume $B\geq0$. The zeta-function is written as
\begin{equation}\label{zeta_H}
    \zeta_{+}(\nu,\omega)=\int_{C}\frac{d\tau \tau^{\nu-1}}{2\pi i}\int d\spx G(\omega,i\tau;\spx,\spx)=V\int_{C}\frac{d\tau \tau^{\nu-1}}{2\pi i} G(\omega,i\tau;\spx,\spx).
\end{equation}
The integrand possesses singularities at the points $i\pi n/B$, $n\in\N$, in the form of simple poles. It also has a cut along the positive part of the real axis. Let us close the contour in the right half-plane. Then
\begin{equation}
    \zeta_{+}(\nu,\omega)=\theta(\omega^2-m^2)\Big(\int_{cut}+\int_{pol}\Big),
\end{equation}
where the shorthand notation was introduced (cf. \cite{ElPerSka94})
\begin{equation}
    \int_{cut}=V\frac{(1-e^{2\pi i(\nu-3/2)})B}{16 \pi^{5/2}}\int_{0}^{\infty}d \tau\frac{\tau^{\nu-3/2}}{\sinh\tau B}e^{-\tau(\omega^2-m^2)},
\end{equation}
and
\begin{equation}
    \int_{pol}=V\frac{e^{i\pi\nu}}{8\pi^{3/2}}\sum_{n=1}^{\infty}(-1)^n\Big(\frac{\pi n}{B}\Big)^{\nu-3/2}
    \Big[e^{i\frac{\pi n}{B}(\omega^2-m^2)+i\frac{\pi}{2}(\nu-3/2)}+c.c.\Big].
\end{equation}
Hereinafter, $c.c.$ denotes a complex conjugate expression constructed in such a way as $\nu$ would be real.

The high-temperature expansion of the partition function for fermions \eqref{expan_ferm} contains the function
\begin{equation}
    \sigma^{l}_{\nu}(\mu)=\int_{0}^{\infty}d \omega (\omega-\mu)^l \zeta_{+}(\nu,\omega)=\sum_{p=0}^{l}C_{l}^{p}(-\mu)^{l-p} \int_{0}^{\infty}d \omega \omega^p \zeta_{+}(\nu,\omega),\quad l=0,1,2,\ldots
\end{equation}
The bosonic expansion \eqref{expan_bos} also includes
\begin{equation}
    \sigma^{-1}_{\nu}(\mu)=\int_{0}^{\infty}d \omega (\omega-\mu)^{-1} \zeta_{+}(\nu,\omega)=\sum_{p=0}^{\infty}\mu^{p} \int_{0}^{\infty}d \omega \omega^{-p-1} \zeta_{+}(\nu,\omega).
\end{equation}
The latter expansion is valid since the integrand vanishes for $|\omega|<\omega_0$, where $\omega_0$ is the minimal particle's energy, and $|\mu|<\omega_0$ for bosons. Thus, we can restrict our consideration to $\sigma^{l}_{\nu}:=\sigma^{l}_{\nu}(\mu=0)$,
\begin{equation}
    \sigma^{l}_{\nu}=\int_{0}^{\infty}d \omega \omega^l \zeta_{+}(\nu,\omega)=\int_{m}^{\infty}d \omega \omega^l [\int_{cut}+\int_{pol}]:=\sigma^{l}_{\text{cut}\ \nu}+\sigma^{l}_{\text{pol}\ \nu},
\end{equation}
where $l$ runs over all integer numbers.

The contribution from the poles is obtained by interchanging the integration and summation order and subsequent integration over $\omega$:
\begin{equation}\label{sigmapol}
    \sigma^{l}_{\text{pol}\ \nu}=V\frac{e^{i \pi \nu}}{16\pi^{3/2}}
    \Big(\frac{B}{\pi}\Big)^{2-\nu+\frac{l}{2}}\Big[ e^{i\frac{\pi}{2}(\nu-1+\frac{l}{2})}\sum_{n=1}^{\infty}\frac{(-1)^n}{n^{2-\nu+l/2}}e^{-i\pi n\bar{m}}\Gamma\Big(\frac{l+1}{2},-i\pi n\bar{m}\Big)+c.c.\Big],
\end{equation}
where it is supposed that $\re\nu>5/2$, and the notation $\bar{m}=m^2/B$ is introduced. As for the contribution from the cut to $\sigma^l_\nu$, having interchanged the integration order, we arrive at
\begin{equation}\label{sigmacut}
    \sigma^{l}_{\text{cut}\ \nu}=V\frac{e^{i \pi \nu}}{16\pi^{5/2}}B^{2-\nu+\frac{l}{2}}\cos\pi\nu\int_{0}^{\infty}d\tau
    \frac{\tau^{\nu-\frac{l}{2}-2}}{\sinh\tau}e^{\tau\bar{m}}\Gamma\Big(\frac{l+1}{2},\tau\bar{m}\Big).
\end{equation}
The integral obtained converges in the domain
\begin{equation}
    \re\nu>3/2\;\;\text{for}\;\;l\leq-1\quad\text{and}\quad\re\nu>l/2+2\;\;\text{for}\;\;l>-1,
\end{equation}
and, in this domain, the change of integration order in $\tau$ and $\omega$ is justified.

The expressions \eqref{sigmapol} and \eqref{sigmacut} can be substantially simplified if one introduces the lower incomplete gamma-function as
\begin{equation}
    \Gamma\Big(\frac{l+1}{2},\tau\bar{m}\Big)=\Gamma\Big(\frac{l+1}{2}\Big)-\gamma\Big(\frac{l+1}{2},\tau\bar{m}\Big).
\end{equation}
The integral over the cut performed with the lower incomplete gamma-function can be represented in the form
\begin{equation}
    \int_{0}^{\infty}d\tau\frac{\tau^{\nu-\frac{l}{2}-2}}{\sinh\tau}e^{\tau\bar{m}}\gamma\Big(\frac{l+1}{2},\tau\bar{m}\Big)=\frac{1}{e^{2\pi i(\nu-\frac{l+1}{2}-\frac{3}{2}+\frac{l+1}{2})}-1}\int_{H}d\tau\frac{\tau^{\nu-\frac{l}{2}-2}}{\sinh\tau}e^{\tau\bar{m}}\gamma\Big(\frac{l+1}{2},\tau\bar{m}\Big),
\end{equation}
where $H$ is the Hankel contour, and recall that $\gamma(s,e^{2\pi i}x)=e^{2\pi i s}\gamma(s,x)$. Then, supposing $\bar{m}<1$, we unfold the contour and bring it to $-\infty$. This results in the contributions from the poles of $1/\sinh \tau$ lying on the imaginary axis. The contributions obtained cancel exactly the part of $\sigma_{\text{pol}}$ containing $\gamma(\frac{l+1}{2},\pm i\pi n \bar{m})$.

As a result, we have
\begin{equation}
    \tilde{\sigma}^{l}_{\text{pol}\ \nu}=V\frac{e^{i \pi \nu}}{16\pi^{3/2}}
    \Big(\frac{B}{\pi}\Big)^{2-\nu+\frac{l}{2}}\Gamma\Big(\frac{l+1}{2}\Big)\Big[e^{i\frac{\pi}{2}(\nu-1+\frac{l}{2})}\sum_{n=1}^{\infty}\frac{(-1)^n}{n^{2-\nu+l/2}}e^{-i\pi n\bar{m}}+c.c.\Big],
\end{equation}
and
\begin{equation}\label{s_cut}
    \tilde{\sigma}^{l}_{\text{cut}\ \nu}=V\frac{e^{i \pi \nu}}{16\pi^{5/2}}B^{2-\nu+\frac{l}{2}}\cos\pi\nu \Gamma\Big(\frac{l+1}{2}\Big)\int_{0}^{\infty}d\tau
    \frac{\tau^{\nu-\frac{l}{2}-2}}{\sinh\tau}e^{\tau\bar{m}}.
\end{equation}
Let us apply the same arguments to the new integrand expression. Now the phase wrap on the lower edge of the cut is $e^{2\pi i(\nu-\frac{l+1}{2}-\frac{3}{2})}$. For odd $l$, it equals to $e^{2\pi i(\nu-\frac{3}{2})}$. In this case, we unfold the contour as above and cancel the contributions. Consequently, $\sigma_{\nu}^{l}=0$ for odd positive $l$. As for odd negative $l$, the considerations above do not work due to singularities of $\Ga((l+1)/2)$.

\paragraph{Strong fields.}

To find the explicit expression for $\tilde{\sigma}_{\text{cut}}$, we develop the exponent in \eqref{s_cut} as a series and change the order of summation and integration. Then the integrals over $\tau$ are reduced to the Riemann zeta function
\begin{equation}
    \tilde{\sigma}^{l}_{\text{cut}\ \nu}=V\frac{e^{i \pi \nu}}{8\pi^{5/2}}B^{2-\nu+\frac{l}{2}}\cos\pi\nu\Gamma\Big(\frac{l+1}{2}\Big)\sum_{k=0}^{\infty} \frac{\bar{m}^{k}}{k!}(1-2^{1-\nu+l/2-k})\Gamma\Big(\nu-\frac{l}{2}-1+k\Big)\zeta\Big(\nu-\frac{l}{2}-1+k\Big).
\end{equation}
In $\tilde{\sigma}_{\text{pol}}$ we also expand the exponent. The summation order in $n$ and $k$ can be interchanged (see, e.g., \cite{Weldon86}). As a result, we deduce
\begin{equation}
    \tilde{\sigma}^{l}_{\text{pol}\ \nu}=V\frac{e^{i \pi \nu}}{8\pi^{3/2}}\Big(\frac{B}{\pi}\Big)^{2-\nu+\frac{l}{2}}\Gamma\Big(\frac{l+1}{2}\Big)\sum_{k=0}^{\infty}\frac{(\pi \bar{m})^k}{k!} \cos\frac{\pi}{2}(\nu-1+l/2-k)(2^{\nu-l/2+k-1}-1)\zeta(l/2+2-k-\nu).
\end{equation}
Making use of the Riemann functional equation for the zeta function in $\tilde{\sigma}_{\text{pol}}$, we join the two contributions
\begin{equation}\label{smunu_bos_ser}
    \sigma_{\nu}^{l}=V\frac{e^{i\pi \nu}}{8\pi^{5/2}}B^{2-\nu+l/2}\cos\frac{\pi l}{2} \Gamma\Big(\frac{l+1}{2}\Big) \sum_{k=0}^{\infty}\frac{(-\bar{m})^k}{k!}(1-2^{1+l/2-\nu-k})\Gamma(\nu-l/2-1+k)\zeta(\nu-l/2-1+k).
\end{equation}
The series obtained converges in the disc $|\bar{m}|<1$, i.e., for the strong fields $|B|>m^2$. Introducing the Hurwitz zeta function \cite{DFGK},
\begin{equation}\label{zeta_hurw}
    \Ga(s)\zeta(s,b+1/2)=\sum_{k=0}^\infty(2^{s+k}-1)\Ga(s+k)\zeta(s+k)\frac{(-b)^k}{k!},\quad |b|<1,\;s\neq1,
\end{equation}
we have
\begin{equation}\label{smunu_bos}
    \sigma_{\nu}^{l}=V\frac{e^{i\pi \nu}}{16\pi^{5/2}}(2B)^{2-\nu+l/2}\cos\frac{\pi l}{2} \Gamma\Big(\frac{l+1}{2}\Big) \Gamma(\nu-l/2-1)\zeta\big(\nu-l/2-1,(1+\bar{m})/2\big).
\end{equation}
This expression vanishes for odd positive $l$.

\paragraph{Weak fields.}

\begin{figure}[t]
\centering
\includegraphics*[width=0.3\linewidth]{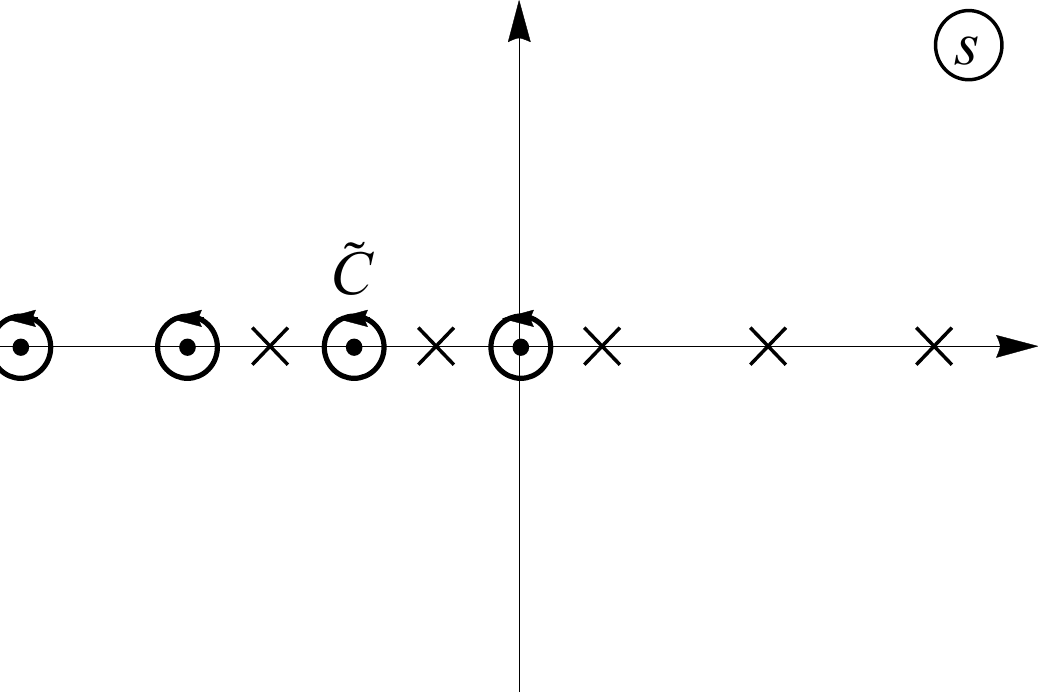}\;
\includegraphics*[width=0.3\linewidth]{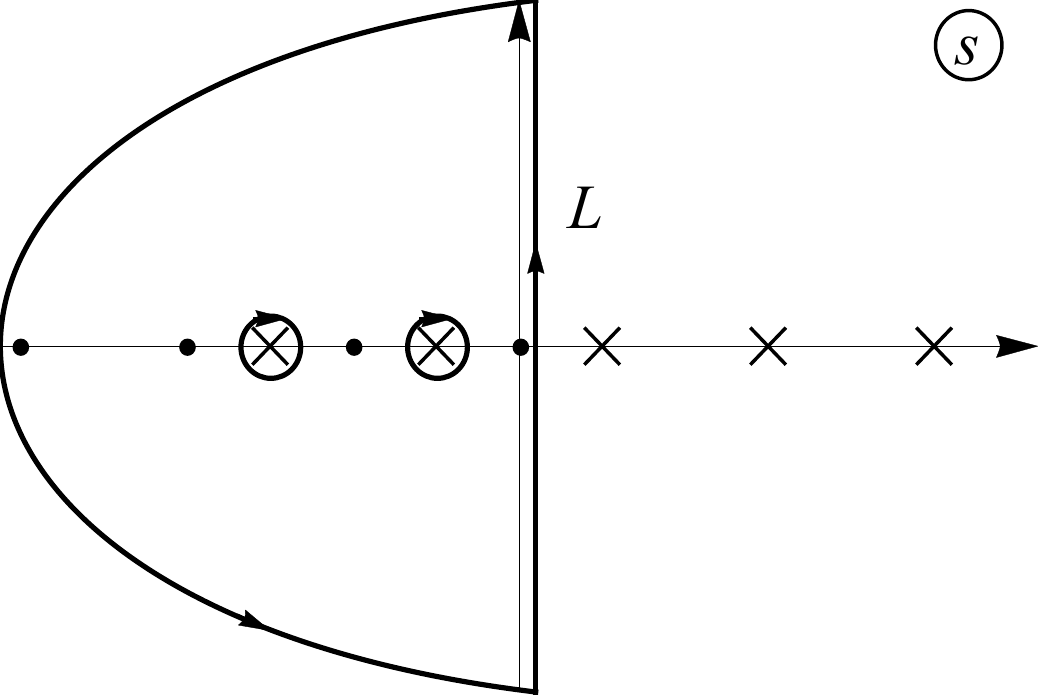}\;
\includegraphics*[width=0.3\linewidth]{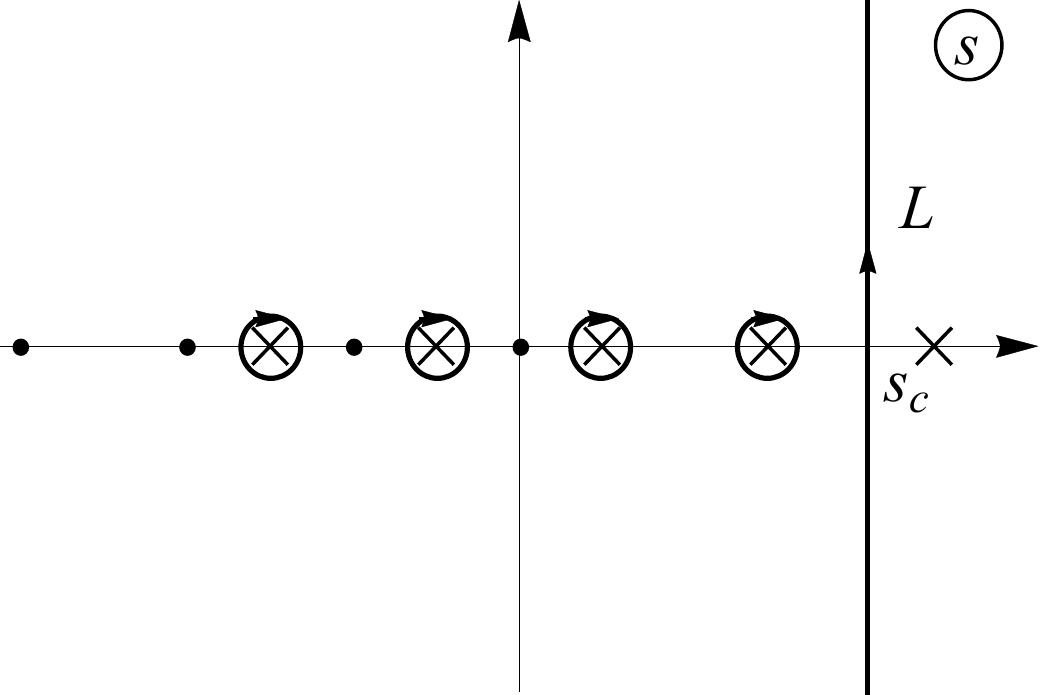}
\caption{{\footnotesize The deformation of the contour $\tilde{C}$ in the $s$ plane (from left to right).}}
\label{contours3}
\end{figure}

In order to find $\sigma_{\nu}^{l}$ in the case of weak fields $|B|<m^2$, it is necessary to resum the series \eqref{smunu_bos_ser}. To this end, we use the Watson method (see, e.g., \cite{Sveshnikov}), rewriting the series in the form of the contour integral
\begin{equation}
    \sum_{k=0}^{\infty}\frac{(-1)^k}{k!} f(k)=\int_{\tilde{C}}\frac{ds}{2\pi i}\Gamma(s)f(-s),
\end{equation}
where the contour $\tilde{C}$ is a union of circles going in the positive direction around the poles of the function $\Gamma(s)$ at the points $s=0,-1,-2,\ldots$, and $f(-s)$ is understood in the sense of analytical continuation. For \eqref{smunu_bos_ser}, the integrand function has poles at the points $s=\nu-l/2+2k, k=-1,0,1,2,\ldots$. Let us blow the contour $\tilde{C}$ (see Fig. \ref{contours3}) so that it intersects the real axis at the point slightly to the right of $s=0$, runs parallel to the imaginary axis, and is closed by the arc of an infinite radius in the left half-plane. The contribution of this arc is zero for $|\bar{m}|<1$. As a result, we have the integral along the contour $L$ converging for $|\arg\bar{m}|<3\pi/2$. Further, we bring $L$ to the right half-plane such that it intersects the real axis at the point $s_{c}$:
\begin{multline}\label{s_munu}
    \sigma_{\nu}^{l}/V=\frac{e^{i\pi \nu}}{8\pi^{5/2}}B^{2-\nu+l/2}\Gamma\Big(\frac{l+1}{2}\Big)\cos\frac{\pi l}{2}\sum_{k=-1}^{k_{max}}\Gamma(\nu-\frac{l}{2}+2k)\bar{m}^{\frac{l}{2}-\nu-2k}(2^{2k+1}-1)\frac{\zeta(-2k-1)}{\Gamma(2k+2)}+\\
    +\frac{e^{i\pi \nu}}{8\pi^{5/2}}B^{2-\nu+l/2}\Gamma\Big(\frac{l+1}{2}\Big)\cos\frac{\pi l}{2}\int_{s_{c}-i\infty}^{s_{c}+i\infty}\frac{ds}{2\pi i}\Gamma(s)\bar{m}^{-s}(1-2^{1+\frac{l}{2}-\nu+s})\Gamma(\nu-\frac{l}{2}-1-s)\zeta(\nu-\frac{l}{2}-1-s),
\end{multline}
where $k_{max}$ is the pole number nearest from the left to $s_{c}$: $k_{max}=[s_c/2+l/4]$ (it is assumed that $s_c$ is large). The first term in \eqref{s_munu} gives the asymptotic expansion of $\s^l_\nu$ with respect to a large mass, while the last term is the remainder of this expansion. The remainder does not have a singularity at $\nu=0$, hence we can safely set $\nu=0$ in it.

We choose $s_{c}$ so as to minimize the remainder and find the explicit expression for it at $\bar{m}$ large. The integrand of \eqref{s_munu} for large $s$ becomes
\begin{equation}\label{asympt}
    \frac{i\sqrt{2}}{4\pi^{l/2+3/2}}\frac{e^{s\ln s-\frac{1}{2}\ln s-s-s\ln(\pi\bar{m})}}{\cos\frac{\pi}{2}(\frac{l}{2}+1+s)}.
\end{equation}
The expression in the exponent possesses the minimum at the point $s_{min}=\bar{m}\pi$. We take the point $s_{c}$ near $s_{min}$ so that it does not hit the poles $s=\nu-l/2+2k$, $k\in \mathbb{Z}$. Such $s_c$ can be always represented as $s_{c}=\bar{m}\pi+\de$, $|\de|<1/2$. Then we parameterize the contour $s=s_{c}+ix$, $x\in\R$, and develop the exponent as a series saving only the terms non-vanishing in the limit $\bar{m}\rightarrow+\infty$. The magnitude of the integration variable $x$ can be taken of the order of unity since the contributions of large values of $x$ are exponentially suppressed due to the cosine in the denominator of \eqref{asympt}. As a result, we obtain
\begin{equation}
    e^{s\ln s-\frac{1}{2}\ln s-s-s\ln(\pi\bar{m})}\approx s_c^{-1/2}e^{s_c\ln\frac{s_{c}}{e\pi\bar{m}}}e^{ix\ln\frac{s_{c}}{\pi\bar{m}}}.
\end{equation}
The resulting integral now has the form
\begin{equation}
    \int_{-\infty}^{\infty}\frac{i dx e^{ix\beta}}{\cos\frac{\pi}{2}(\alpha-ix)}= 2i (-1)^k \frac{e^{\beta\alpha_0}}{\cosh\beta}, \quad \alpha_0\in (-1,1), \quad k\in \mathbb{Z}, \quad \alpha=:\alpha_0+2k.
\end{equation}
Then the last term in \eqref{s_munu} becomes
\begin{equation}\label{nonpert_bos}
    \sigma_{\text{np}\ \nu}^{l}/V=\frac{(-1)^{[(\pi\bar{m}+l/2)/2]}}{8\pi^{5/2}}\Big(\frac{B}{\pi}\Big)^{2+l/2}\cos\frac{\pi l}{2}\Ga\Big(\frac{l+1}{2}\Big)\frac{e^{-\pi\bar{m}}}{\sqrt{2\bar{m}}},
\end{equation}
where we neglect all the subleading contributions in the limit $\bar{m}\rightarrow+\infty$. Obviously, this contribution is non-perturbative, i.e., it cannot be reproduced by a naive summation of the Feynman diagrams and is suppressed by the exponential factor $e^{-\pi m^2/B}$.

\paragraph{Coefficients $\zeta_k(\nu)$.}

Apart from the functions $\sigma_{\nu}^l$, the high-temperature expansion also contains the coefficients $\zeta_k(\nu)$ \eqref{zeta_knu}, which are obtained if one represents the trace of the heat kernel in $\zeta_+(\nu,\omega)$ in the form of the standard heat kernel expansion and formally integrates the series over $\tau$ termwise. Then we have for \eqref{HK_bos}
\begin{equation}
    \frac{B\tau^{-1/2}}{\sinh \tau B}=\sum_{k=0}^{\infty}(2^k-2)\frac{\zeta(1-k)}{\Gamma(k)}B^k \tau^{k-3/2}=:\sum_{k=0}^{\infty}a_k \tau^{k-3/2},
\end{equation}
where $a_k$ are the coefficients of the heat kernel expansion. Substituting this expansion into \eqref{zeta_H} and integrating the series over $\tau$, we deduce from the definition \eqref{zeta_knu} for the non-vanishing coefficients
\begin{equation}
    \zeta_{2k+2s}(\nu)=Ve^{i\pi\nu}\frac{(-1)^{k+s}}{(4\pi)^{3/2}}\frac{a_km^{2s}}{s!\Gamma(5/2-\nu-k-s)},
\end{equation}
where $k$ and $s$ run over all the natural numbers and zero.

\subsection{Dirac fermions}

Let us briefly consider the high-temperature expansion of the one-loop contribution of the Dirac fermions to the omega-potential. The trace over spinor indices of the diagonal of the heat kernel for Dirac fermions takes the form \cite{ElmSkag95}
\begin{equation}
    G(\omega,i\tau;\spx,\spx)=\frac{e^{3\pi i/2}}{(4 \pi)^{3/2}\tau^{1/2}}2B\coth\tau B e^{-\tau(\omega^2-m^2)}.
\end{equation}
The analogues of  \eqref{sigmapol}, \eqref{sigmacut} in $\sigma^l_\nu$ are written as
\begin{equation}
\begin{split}
    \sigma^{l}_{\text{pol}\ \nu}&=V\frac{e^{i \pi \nu}}{8\pi^{3/2}}
    \Big(\frac{B}{\pi}\Big)^{2-\nu+\frac{l}{2}}\Big[e^{i\frac{\pi}{2}(\nu-1+\frac{l}{2})}\sum_{n=1}^{\infty}\frac{e^{-i\pi n\bar{m}}}{n^{2-\nu+l/2}}\Gamma\Big(\frac{l+1}{2},-i\pi n\bar{m}\Big)+c.c.\Big],\\
    \sigma^{l}_{\text{cut}\ \nu}&=V\frac{e^{i \pi \nu}}{8\pi^{5/2}}B^{2-\nu+\frac{l}{2}}\cos\pi\nu\int_{0}^{\infty}d\tau
    \tau^{\nu-\frac{l}{2}-2}(\coth\tau-1) e^{\tau\bar{m}}\Gamma\Big(\frac{l+1}{2},\tau\bar{m}\Big)+\\
    &+V\frac{e^{i\pi \nu}}{4\pi^{5/2}} B \cos\pi\nu \int_{m}^{\infty}d\omega \omega^l\int_{0}^{\infty}d\tau\tau^{\nu-3/2}e^{-\tau(\omega^2-m^2)}.
\end{split}
\end{equation}
Further, the considerations are completely analogous to the given above in the bosonic case. The difference is only that, in unfolding the contour, one needs to demand  $\bar{m}<2$. As a result, we deduce
\begin{equation}
\begin{split}
    \tilde{\sigma}^{l}_{\text{pol}\ \nu}&=V\frac{e^{i \pi \nu}}{8\pi^{3/2}}
    \Big(\frac{B}{\pi}\Big)^{2-\nu+\frac{l}{2}}\Gamma\Big(\frac{l+1}{2}\Big)\Big[e^{i\frac{\pi}{2}(\nu-1+\frac{l}{2})}\sum_{n=1}^{\infty}\frac{e^{-i\pi n\bar{m}}}{n^{2-\nu+l/2}}+c.c.\Big],\\
    \tilde{\sigma}^{l}_{\text{cut}\ \nu}&=V\frac{e^{i \pi \nu}}{8\pi^{5/2}}B^{2-\nu+\frac{l}{2}}\cos\pi\nu \Gamma\Big(\frac{l+1}{2}\Big)\int_{0}^{\infty}d\tau
    \tau^{\nu-\frac{l}{2}-2}(\coth\tau-1)e^{\tau\bar{m}}-\\
    &-V\frac{e^{i\pi \nu}}{8\pi^{5/2}}m^{2-2\nu+l}B \cos\frac{\pi l}{2} \Gamma\Big(\frac{l+1}{2}\Big)\Gamma\Big(\nu-\frac{l}{2}-1\Big).
\end{split}
\end{equation}

\paragraph{Strong fields.}

Substituting
\begin{equation}
    \coth\tau-1=\frac{2}{e^{2\tau}-1}
\end{equation}
into $\tilde{\sigma}_{\text{cut}}$ and expanding the exponent in the integrand, we obtain
\begin{equation}
\begin{split}
    \tilde{\sigma}^{l}_{\text{cut}\ \nu}&=V\frac{e^{i \pi \nu}}{4\pi^{5/2}}B^{2-\nu+\frac{l}{2}}\cos\pi\nu\Gamma\Big(\frac{l+1}{2}\Big)\sum_{k=0}^{\infty}\frac{\bar{m}^{k}}{k!}2^{1-\nu+l/2-k}\Gamma\Big(\nu-\frac{l}{2}-1+k\Big) \zeta\Big(\nu-\frac{l}{2}-1+k\Big)-\\
    &-V\frac{e^{i\pi \nu}}{8\pi^{5/2}}m^{2-2\nu+l}B \cos\frac{\pi l}{2} \Gamma\Big(\frac{l+1}{2}\Big)\Gamma\Big(\nu-\frac{l}{2}-1\Big).
\end{split}
\end{equation}
Let us perform the same procedure with the function $\tilde{\sigma}_{\text{pol}}$. The order of summation over $n$ and $k$ can be interchanged according to the formula (see, e.g., \cite{Weldon86})
\begin{equation}
    \sum_{n=1}^{\infty}\frac{e^{ian}}{n^{s}}=\sum_{k=0}^{\infty}\zeta(s-k)\frac{(ia)^k}{k!}-\Gamma(1-s)(ia)^{s-1}e^{-i\pi s},\quad\re s<0,\;a\in \mathbb{R}.
\end{equation}
Then, using the Riemann functional equation for the zeta function, we have
\begin{equation}
\begin{split}
    \tilde{\sigma}_{\text{pol}\ \nu}^{l} &= V \frac{e^{i\pi\nu}}{4\pi^{5/2}}B^{2-\nu+l/2}\Gamma\Big(\frac{l+1}{2}\Big)\sum_{k=0}^{\infty}\frac{\bar{m}^k}{k!}2^{1-\nu+l/2-k}\Gamma(\nu-l/2-1+k)\zeta(\nu-l/2-1+k)\times\\ &\times[(-1)^k\cos\frac{\pi l}{2}-\cos\pi\nu]
    +V\frac{e^{i\pi\nu}}{4\pi^{5/2}}m^{2-2\nu+l}B\cos\frac{\pi l}{2}\Gamma\Big(\frac{l+1}{2}\Big)\Gamma(\nu-l/2-1).
\end{split}
\end{equation}
As a result, adding the two contributions $\tilde{\s}_{\text{cut}}$ and $\tilde{\s}_{\text{pol}}$, we come to
\begin{equation}\label{s_munu_ferm}
\begin{split}
    \sigma_{\nu}^{l}&=V\frac{e^{i\pi\nu}}{4\pi^{5/2}}B^{2-\nu+l/2} \cos\frac{\pi l}{2} \Gamma\Big(\frac{l+1}{2}\Big) \sum_{k=0}^{\infty}\frac{(-\bar{m})^k}{k!}2^{1-\nu+l/2-k}\Gamma(\nu-l/2-1+k)\zeta(\nu-l/2-1+k)+\\
    &+V\frac{e^{i\pi\nu}}{8\pi^{5/2}}m^{l+2-2\nu}B\cos\frac{\pi l}{2}\Gamma\Big(\frac{l+1}{2}\Big)\Gamma(\nu-l/2-1).
\end{split}
\end{equation}
The series obtained is convergent in the domain $|\bar{m}|<2$. Introducing the Hurwitz zeta function \cite{NIST},
\begin{equation}
    \Gamma(s)\zeta(s,a)=\sum_{k=0}^\infty\Gamma(s+k)\zeta(s+k)\frac{(1-a)^k}{k!},\quad |1-a|<1,\;s\neq1,
\end{equation}
we arrive at
\begin{equation}\label{smunu_ferm}
    \sigma_{\nu}^{l}=V\frac{e^{i\pi\nu}}{8\pi^{5/2}} \cos\frac{\pi l}{2} \Gamma\Big(\frac{l+1}{2}\Big)\Gamma(\nu-l/2-1)\Big[ (2B)^{2-\nu+l/2}\zeta(\nu-\l/2-1,1+\bar{m}/2)+ m^{2-2\nu+l}B\Big].
\end{equation}
The expression \eqref{smunu_bos} in the previous section can be obtained from \eqref{smunu_ferm} if one observes that
\begin{equation}
    \frac{1}{\sinh\tau}=\coth\frac{\tau}{2}-\coth\tau,
\end{equation}
and substitutes this representation into \eqref{HK_bos}. This gives an indirect check of the both expressions.

\paragraph{Weak fields.}

The consideration is completely equivalent to the consideration in the previous section with the exception that there arises an additional pole of the gamma function at the point $s=\nu-l/2-1$ with the residue
\begin{equation}
    -V\frac{e^{i\pi\nu}}{8\pi^{5/2}}B^{2-\nu+l/2}\bar{m}^{l/2-\nu+1}\cos\frac{\pi l}{2}\Gamma\Big(\frac{l+1}{2}\Big)\Gamma(\nu-l/2-1).
\end{equation}
The contribution of this pole cancels the second term in \eqref{s_munu_ferm}. Thus we have the expansion
\begin{multline}\label{s_munu_dir}
    \sigma_{\nu}^{l}/V=-\frac{e^{i\pi \nu}}{4\pi^{5/2}}B^{2-\nu+l/2}\cos\frac{\pi l}{2}\Gamma\Big(\frac{l+1}{2}\Big) \sum_{k=-1}^{k_{max}}\Gamma\Big(\nu-\frac{l}{2}+2k\Big)\bar{m}^{\frac{l}{2}-\nu-2k}2^{2k+1}\frac{\zeta(-2k-1)}{\Gamma(2k+2)}+\\
    +\frac{e^{i\pi \nu}}{4\pi^{5/2}}B^{2-\nu+l/2}\cos\frac{\pi l}{2} \Gamma\Big(\frac{l+1}{2}\Big) \int_{s_{c}-i\infty}^{s_{c}+i\infty}\frac{ds}{2\pi i}\Gamma(s)\bar{m}^{-s}2^{1+\frac{l}{2}-\nu+s}\Gamma\Big(\nu-\frac{l}{2}-1-s\Big)\zeta\Big(\nu-\frac{l}{2}-1-s\Big).
\end{multline}
For large $s$, $\re s>0$, the last term in this expression differs from the analogous term in \eqref{s_munu} only by the factor $-2$. Therefore, at $\bar{m}\rightarrow+\infty$, the last term in \eqref{s_munu_dir} becomes
\begin{equation}
    \sigma_{\text{np}\ \nu}^{l}/V=-\frac{(-1)^{[(\pi\bar{m}+l/2)/2]}}{4\pi^{5/2}}\Big(\frac{B}{\pi}\Big)^{2+l/2}\cos\frac{\pi l}{2}\Ga\Big(\frac{l+1}{2}\Big)\frac{e^{-\pi\bar{m}}}{\sqrt{2\bar{m}}}.
\end{equation}
This contribution is essentially non-perturbative.

\paragraph{Coefficients $\zeta_k(\nu)$.}

Inasmuch as
\begin{equation}
    2B\tau^{-1/2}\coth \tau B=-\sum_{k=0, \neq1}^{\infty}2^{k+1}\frac{\zeta(1-k)}{\Gamma(k)}B^k \tau^{k-3/2}=:\sum_{k=0}^{\infty}a_k \tau^{k-3/2},
\end{equation}
the non-vanishing coefficients
\begin{equation}
    \zeta_{2k+2s}(\nu)=Ve^{i\pi\nu}\frac{(-1)^{k+s}}{(4\pi)^{3/2}}\frac{a_km^{2s}}{s!\Gamma(5/2-\nu-k-s)},
\end{equation}
where $k$ and $s$ run over all natural numbers and zero.

\section{High-temperature expansions}\label{High-Temp_Expan}
\subsection{Scalar particles}

As follows from the spin-statistics theorem, the scalar particles are bosons. So, using the bosonic expansion \eqref{expan_bos} and canceling the poles in the complex $\nu$ plane, we derive the complete (up to the exponentially suppressed at $\be\rightarrow+0$ terms) high-temperature expansion in the case of strong fields
\begin{multline}\label{omega_b}
    -\frac{\Omega_b(\mu)}{V}=\sideset{}{'}\sum_{k,s,n=0}^{\infty} (-1)^{k+s}a_km^{2s}\frac{\Gamma(4-2k-2s)\zeta(4-2k-2s-n)}{(4\pi)^{3/2}s!\Gamma(\frac{5}{2}-k-s)\beta^{4-2k-2s}}\frac{(\beta\mu)^n}{n!}+\\
    +\sideset{}{'}\sum_{k,p=0}^{\infty}\frac{\beta^{-1}B^{\frac{3}{2}-\frac{p}{2}}\mu^p}{(4\pi)^{3/2}\Gamma(\tfrac{p}{2}+1)k!}\Big(-\frac{m^2}{B}\Big)^k (1-2^{\frac{1}{2}-\frac{p}{2}-k})\Gamma(\tfrac{p}{2}-\tfrac{1}{2}+k)\zeta(\tfrac{p}{2}-\tfrac{1}{2}+k)+\frac{\mu m^2}{8\pi^2\be}\Big[\big(1 -\tfrac23\tfrac{\mu^2}{m^2}\big)\ln\tfrac{\be^2 B}{8e^\ga}+\tfrac23\tfrac{\mu^2}{m^2}\Big]+\\
    +\sideset{}{'}\sum_{l,p,k=0}^{\infty}(\mu\beta)^l \frac{(-1)^p \zeta(-l)}{p!(l-p)!} \frac{B^{2+\frac{p}{2}}\mu^{-p}}{(4\pi)^{3/2}\Gamma(\frac{1-p}{2})k!} \Big(-\frac{m^2}{B}\Big)^k (1-2^{1+\frac{p}{2}-k})\Gamma(k-1-\tfrac{p}{2})\zeta(k-1-\tfrac{p}{2})-\\
    -\sideset{}{'}\sum_{\substack{k,s,n=0\\2+s\geq k}}^{\infty}\frac{(-1)^{s} \zeta(-2s-n)\beta^{2s+n}\mu^n m^{4-2k+2s}a_k}{16\pi^{3/2}(2s)!n!\Gamma(\frac{1}{2}-s)(2-k+s)!}
    \Big[\ln(\be^2B)-\frac{\zeta'(1-k)}{\zeta(1-k)}-\psi(k)-\frac{2^{k}\ln 2}{2-2^k}+\\
    +\psi\big(\tfrac{1}{2}-s\big)-2\frac{\zeta'(-2s-n)}{\zeta(-2s-n)}-2\psi(2s+1)\Big],
\end{multline}
where the primes mean that all singular terms of the series should be discarded. The last term in \eqref{omega_b} contains by definition only the odd powers of $\mu$ save the term with $n=s=0$. Therefore, the finite and divergent at $\beta\rightarrow0$ parts of the expansion look as follows
\begin{multline}\label{omega_bs}
    \frac{\Omega_b(\mu)}{V}=-\frac{\pi^2}{90} \be^{-4}-\frac{\zeta(3)}{\pi^2} \mu \be^{-3}-\Big(\frac{\mu^2}{12}-\frac{m^2}{24}\Big)\be^{-2} -\\
    -\sideset{}{'}\sum_{k,p=0}^{\infty}\frac{\beta^{-1}B^{\frac{3}{2}-\frac{p}{2}}\mu^p}{(4\pi)^{3/2}\Gamma(\tfrac{p}{2}+1)k!}\Big(-\frac{m^2}{B}\Big)^k (1-2^{\frac{1}{2}-\frac{p}{2}-k})\Gamma(\tfrac{p}{2}-\tfrac{1}{2}+k)\zeta(\tfrac{p}{2}-\tfrac{1}{2}+k)-\frac{\mu m^2}{8\pi^2\be}\Big[\big(1 -\tfrac23\tfrac{\mu^2}{m^2}\big)\ln\tfrac{\be^2 B}{8e^\ga}+\tfrac23\tfrac{\mu^2}{m^2}\Big]\\
    +\frac{\mu^4}{48\pi^2}-\frac{\mu^2 m^2}{16\pi^2} -\Big(\frac{m^4}{64\pi^2}-\frac{B^2}{192\pi^2}\Big)\ln\tfrac{\be^2Be^\ga}{32\pi^2}-\frac{B^2}{192\pi^2}\Big(\frac{\zeta'(-1)}{\zeta(-1)}-\ln\tfrac{8}{e^{1-\ga}}\Big) +\frac{B^2}{16\pi^2}\sideset{}{'}\sum_{k=0}^{\infty} \Big(-\frac{m^2}{B}\Big)^k\frac{(1-2^{1-k})\zeta(k-1)}{k(k-1)}.
\end{multline}
The complete expansion in the case of weak fields reads as
\begin{multline}
    -\frac{\Omega_b(\mu)}{V}=\sideset{}{'}\sum_{k,s,n=0}^{\infty} (-1)^{k+s}a_km^{2s}\frac{\Gamma(4-2k-2s)\zeta(4-2k-2s-n)}{(4\pi)^{3/2}s!\Gamma(\frac{5}{2}-k-s)\beta^{4-2k-2s}}\frac{(\beta\mu)^n}{n!}+\\
    +\beta^{-1}\sum^{\infty}_{p=0}\sideset{}{'}\sum_{k=-1}^{k_{max}}\frac{B^{\frac{3}{2}-\frac{p}{2}}\mu^p }{(4\pi)^{3/2}}\frac{\Gamma(\frac{p}{2}+\frac{1}{2}+2k)}{\Gamma(\frac{p}{2}+1)} \Big(\frac{m^2}{B}\Big)^{-\frac{p}{2}-\frac{1}{2}-2k}(2^{2k+1}-1)\frac{\zeta(-2k-1)}{\Gamma(2k+2)} +\frac{\mu m^2}{8\pi^2\be}\big(1-\tfrac{2}{3}\tfrac{\mu^2}{m^2}\big)\ln\tfrac{\be^2m^2}{4e}+\\
    +\sum_{l,p=0}^{\infty}\sideset{}{'}\sum_{k=-1}^{k_{max}}(\mu\be)^l\frac{(-1)^p \zeta(-l)}{p!(l-p)!} \frac{B^{2+\frac{p}{2}}\mu^{-p}\Gamma(2k-\frac{p}{2})}{(4\pi)^{3/2}\Gamma(\frac{1-p}{2})}\Big(\frac{B}{m^2}\Big)^{2k-\frac{p}{2}}(2^{2k+1}-1)\frac{\zeta(-2k-1)}{\Gamma(2k+2)}-\\
    -\sideset{}{'}\sum_{\substack{k,s,n=0\\2+s\geq k}}^{\infty}\frac{(-1)^{s}\zeta(-2s-n)\beta^{2s+n}\mu^n m^{4-2k+2s}a_k}{16\pi^{3/2}(2s)!n!\Gamma(\tfrac{1}{2}-s)(2-k+s)!} \Big[\ln(\beta^2m^2)-\psi(3-k+s)
    +\psi(\tfrac{1}{2}-s)\\
    -2\frac{\zeta'(-2s-n)}{\zeta(-2s-n)}-2\psi(2s+1)\Big]+\dots,
\end{multline}
where the dots denote the exponentially suppressed terms coming from \eqref{nonpert_bos}. In this formula, the terms are discarded by the same rule as in \eqref{omega_b}.

The sum over $p$ in the second term can be expressed through the hypergeometric function. The finite and divergent at $\beta\rightarrow0$ part of the expansion takes the form
\begin{multline}\label{omega_b_apr}
    \frac{\Omega_b(\mu)}{V}=-\frac{\pi^2}{90}\be^{-4}-\frac{\zeta(3)}{\pi^2} \mu\be^{-3}-\Big(\frac{\mu^2}{12}-\frac{m^2}{24}\Big)\be^{-2}-\frac{\mu m^2}{8\pi^2\be}\big(1-\tfrac{2}{3}\tfrac{\mu^2}{m^2}\big)\ln\tfrac{\be^2m^2}{4e}+\frac{\mu^4}{48\pi^2}-\frac{\mu^2 m^2}{16\pi^2} -\\
    -\sideset{}{'}\sum_{k=-1}^{k_{max}}\frac{\be^{-1}m^3}{(4\pi)^{3/2}}\Big(\frac{B}{m^2}\Big)^{2k+2}(2^{2k+1}-1) \frac{\zeta(-2k-1)}{\Gamma(2k+2)} \bigg[\frac{\Gamma(2k+\tfrac{1}{2})}{\big(1-\tfrac{\mu^2}{m^2}\big)^{2k+\frac{1}{2}}} +\frac{2\mu}{\pi^{1/2}m}\Gamma(2k+1)F(1,2k+1;\tfrac{3}{2};\tfrac{\mu^2}{m^2})\bigg]-\\
    -\Big(\frac{m^4}{64\pi^2}-\frac{B^2}{192\pi^2}\Big)\ln\tfrac{\be^2m^2e^{2\ga}}{16\pi^2}+\frac{3m^4}{128\pi^2} +\frac{B^2}{16\pi^2}\sum^{k_{max}}_{k=1}\Big(\frac{B}{m^2}\Big)^{2k}\frac{(2^{2k+1}-1)\zeta(-2k-1)}{2k(2k+1)}+\cdots
\end{multline}
The expansion derived in \cite{DFGK} readily follows from the above expansions if one takes into account the contribution of antiparticles to the omega-potential, viz., if one adds the same expression with $\mu\rightarrow-\mu$. The high-temperature expansion found in \cite{DFGK} is an even in $\mu$ part of the expansion \eqref{omega_b}. In particular, formulas \eqref{omega_bs}, \eqref{omega_b_apr} allow one to find separately the number of particles and antiparticles in the system \cite{KhalilB} as
\begin{equation}
    N=-\frac{\partial\Omega_b}{\partial\mu},
\end{equation}
and not only the total charge. For example, it follows from \eqref{omega_bs} at $\mu=0$ (zero total charge) that
\begin{equation}
    \frac{N}{V}=\frac{\zeta(3)}{\pi^2}\be^{-3}+\frac{\be^{-1}}{8\pi^2}\Big[B\ln\frac{\Ga^2(1+m^2/B)}{2\Ga^2(1+m^2/2B)}+m^2\ln\frac{\be^2B}{8}\Big],
\end{equation}
where, as in \eqref{omega_bs}, the terms vanishing in the limit $\be\rightarrow0$ are cast out.

\subsection{Scalar fermions}

Let us derive the explicit expressions for the fermionic expansion \eqref{expan_ferm} for scalars. This expansion can be used to obtain the energy of zero-point fluctuations \eqref{eff_act_1b}. From \eqref{expan_ferm}, canceling the poles in the $\nu$ plane, we derive in the case of strong fields
\begin{equation}\label{omega_f}
\begin{split}
    -\frac{\Omega_f(\mu)}{V}&=\sideset{}{'}\sum_{k,s,n=0}(-1)^{k+s}a_k m^{2s} \frac{\Gamma(4-2k-2s)\eta(4-2k-2s-n)}{(4\pi)^{3/2}s!\Gamma(\frac{5}{2}-k-s)\beta^{4-2k-2s}} \frac{(\beta\mu)^n}{n!}+\\
    &+\sideset{}{'}\sum_{l,p,k=0} (\beta\mu)^lB^2\Big(\frac{B}{\mu^2}\Big)^{\frac{p}{2}}\Big(-\frac{m^2}{B}\Big)^{k} \frac{\eta(-l)(1-2^{1+p/2-k})\Gamma(k-p/2-1)\zeta(k-p/2-1)}{(4\pi)^{3/2}p!(l-p)!k!\Gamma(\frac{1-p}{2})} -\\
    &-\sideset{}{'}\sum_{\substack{k,r,n=0\\2+s\geq k}} \frac{(-1)^{s}\eta(-2s-n)(\beta\mu)^n (\beta m)^{2s}}{16\pi^{3/2}(2s)!n!(2-k+s)!\Gamma(\tfrac{1}{2}-s)}m^{4-2k}a_k
    \Big[\ln(\beta^2B)-\psi(k)-\frac{\zeta'(1-k)}{\zeta(1-k)}-\\
    &-\frac{2^k \ln 2}{2-2^k}+\psi(\tfrac{1}{2}-s)-2\frac{\zeta'(-2s-n)}{\zeta(-2s-n)}-\frac{4 \ln 2}{2^{-2s-n}-2}-2\psi(2s+1)\Big],
\end{split}
\end{equation}
where $\eta(s):=(1-2^{1-s})\zeta(s)$ is the Dirichlet $\eta$ function. The primes at the sums mean the same as in \eqref{omega_b}. The last term contains by definition only the odd powers of $\mu$ except the term with $n=s=0$. Explicitly, the finite and divergent at $\beta\rightarrow0$ part of the high-temperature expansion reads as
\begin{multline}
    \frac{\Omega_{f}(\mu)}{V}=-\frac{7\pi^2}{720}\be^{-4}-\frac{3\zeta(3)}{4\pi^2}\mu \be^{-3}+\Big(\frac{m^2}{48}-\frac{\mu^2}{24}\Big)\be^{-2}+\frac{\mu\ln2}{12\pi^2}(3m^2-2\mu^2)\be^{-1}+\frac{m^2\mu^2}{16\pi^2}-\frac{\mu^4}{48\pi^2}-\\
    -\frac{B^2}{16\pi^2}\sum^{\infty}_{k=1,\neq2}\frac{(-1)^k}{k(k-1)}\Big(\frac{m^2}{B}\Big)^k(1-2^{1-k})\zeta(k-1) +\Big(\frac{m^4}{64\pi^2}-\frac{B^2}{192\pi^2}\Big) \ln\tfrac{\be^2Be^\ga}{2\pi^2}+\frac{B^2}{192\pi^2}\Big(\frac{\zeta'(-1)}{\zeta(-1)}-\ln\tfrac{8}{e^{\ga-1}}\Big).
\end{multline}
According to \eqref{eff_act_1b}, the non-renormalized energy of vacuum fluctuations of charged scalar bosons in a strong magnetic field takes the form
\begin{multline}
    E_{vac}=2\partial_{\beta_{0}}(\beta_{0} \Omega_{f}(0))=V\Big[\frac{7\pi^2}{120}\beta_0^{-4}-\frac{m^2}{24}\beta_0^{-2}+\Big(\frac{m^4}{16\pi^2}-\frac{B^2}{48\pi^2}\Big)\ln(e \beta_0)+\\
    -\frac{B^2}{8\pi^2}\sum_{k=1,\neq2}^{\infty}\frac{(1-2^{1-k})\zeta(k-1)}{k(k-1)}\Big(-\frac{m^2}{B}\Big)^k +\Big(\frac{m^4}{32\pi^2}-\frac{B^2}{96\pi^2}\Big)\ln\frac{Be^\ga}{2\pi^2} +\frac{B^2}{96\pi^2}\Big(\frac{\zeta'(-1)}{\zeta(-1)}-\ln\frac{8}{e^{\ga-1}}\Big)\Big].
\end{multline}
Here $\beta_{0}$ is to be understood as some cut-off parameter.

In the case of weak fields, we have
\begin{equation}
\begin{split}
    -\frac{\Omega_{f}(\mu)}{V}&=\sideset{}{'}\sum_{k,s,n=0}^{\infty}\Gamma(4-2k-2s)\eta(4-2k-2s-n)\beta^{-(4-2k-2s)}\frac{(\beta\mu)^n}{n!}\frac{(-1)^{k+s}a_k}{(4\pi)^{3/2}} \frac{m^{2s}}{s!\Gamma(\frac{5}{2}-k-s)}+\\
    &+\sum_{l,s=0}^{\infty}\sideset{}{'}\sum_{k=-1}^{k_{max}}\frac{B^2}{(4\pi)^{3/2}}\frac{\eta(-l)}{(2s)!(l-2s)!}\frac{\Gamma(2k-s)}{\Gamma(\tfrac{1}{2}-s)}\Big(\frac{m}{\mu}\Big)^{2s} (\beta\mu)^l \Big(\frac{B}{m^2}\Big)^{2k}(2^{2k+1}-1)\frac{\zeta(-2k-1)}{\Gamma(2k+2)}-\\
    &-\sideset{}{'}\sum^{\infty}_{\substack{k,s,n=0 \\2+s\geq k}} \frac{(-1)^{s}\eta(-2s-n)(\beta\mu)^n (\beta m)^{2s}}{16\pi^{3/2}(2s)!n!(2-k+s)!\Gamma(\tfrac{1}{2}-s)}m^{4-2k}a_k \times\\
    &\times\Big[\ln (\beta^2m^2)-\psi(3-k+s)+\psi(\tfrac{1}{2}-s)-2\frac{\zeta'(-2s-n)}{\zeta(-2s-n)}-\frac{4\ln2}{2^{-n-2s}-2}-2\psi(2s+1)\Big].
\end{split}
\end{equation}
Here the notation and conventions are the same as in \eqref{omega_b}, \eqref{omega_f}. The exponentially suppressed contributions are thrown away. Then the finite and divergent parts can be cast into the form
\begin{multline}
    \frac{\Omega_{f}(\mu)}{V}=-\frac{7\pi^2}{720}\be^{-4}-\frac{3\zeta(3)}{4\pi^2}\mu \be^{-3}+\frac{1}{48}(m^2-2\mu^2)\be^{-2}+\frac{\mu\ln2}{12\pi^2}(3m^2-2\mu^2)\be^{-1}+\frac{m^2\mu^2}{16\pi^2}-\frac{\mu^4}{48\pi^2}-\\
    -\frac{B^2}{16\pi^2}\sum^{k_{max}}_{k=1}\frac{(2^{2k+1}-1)\zeta(-2k-1)}{2k(2k+1)}\Big(\frac{B}{m^2}\Big)^{2k}
    +\Big(\frac{m^4}{32\pi^2}-\frac{B^2}{96\pi^2}\Big)\ln\frac{\be me^\ga}{\pi}-\frac{3m^4}{128\pi^2}.
\end{multline}
Consequently, the non-renormalized energy of vacuum fluctuations of charged scalar bosons in weak fields is written as
\begin{multline}
    E_{vac}=2\partial_{\beta_{0}}(\beta_{0} \Omega_{f}(0))=V\Big[\frac{7\pi^2}{120}\beta_0^{-4}-\frac{m^2}{24}\beta_0^{-2}+\Big(\frac{m^4}{16\pi^2}-\frac{B^2}{48\pi^2}\Big)\ln\frac{ \beta_0me^{\ga+1}}{\pi}-\frac{3m^4}{64\pi^2}-\\
    -\frac{B^2}{8\pi^2}\sum^{k_{max}}_{k=1}\frac{(2^{2k+1}-1)\zeta(-2k-1)}{2k(2k+1)}\Big(\frac{B}{m^2}\Big)^{2k} -\frac{(-1)^{[\pi m^2/2B]}}{8\pi^2}\Big(\frac{B}{\pi}\Big)^2 \Big(\frac{B}{2m^2}\Big)^{1/2}e^{-\pi m^2/B}\Big].
\end{multline}
The renormalization of the one-loop contribution is performed in the standard way (see, e.g., \cite{BogShir}). The counterterms are added to the initial action of the theory. They have the mass dimension less than or equal to 4 (without taking into account the dimension of $\be_0$), must cancel all the divergencies, and set the coupling constants to their physical values. In our case, the counterterms that should be added to the initial Lagrangian have the form
\begin{equation}\label{counterterms}
    c.t.=\frac{7\pi^2}{120}\beta_0^{-4}-\frac{m^2}{24}\beta_0^{-2}+\Big(\frac{m^4}{16\pi^2}-\frac{B^2}{48\pi^2}\Big)\ln\frac{ \beta_0me^{\ga+1}}{\pi}-\frac{3m^4}{64\pi^2}.
\end{equation}
This corresponds to the choice of the value of the fine structure constant $\al$ that is observed in low-energy experiments in the absence of the external fields. As a result, the renormalized vacuum contribution in the limit of weak fields is
\begin{equation}
    \frac{E^{ren}_{vac}}{V}=\frac{E_{vac}}{V}-c.t.=-\frac{B^2}{8\pi^2}\sum^{k_{max}}_{k=1}\frac{(2^{2k+1}-1)\zeta(-2k-1)}{2k(2k+1)}\Big(\frac{B}{m^2}\Big)^{2k}-\frac{(-1)^{[\pi m^2/2B]}}{8\pi^2}\Big(\frac{B}{\pi}\Big)^2 \Big(\frac{B}{2m^2}\Big)^{1/2}e^{-\pi m^2/B},
\end{equation}
which coincides exactly with formula (1.34) of \cite{Dunne2} for the effective Lagrangian without the exponentially suppressed contribution. In the case of strong fields, we have
\begin{multline}
    \frac{E^{ren}_{vac}}{V}=\frac{E_{vac}}{V}-c.t.=-\frac{B^2}{8\pi^2}\sum^{\infty}_{k=3}\frac{(1-2^{1-k})\zeta(k-1)}{k(k-1)}\Big(-\frac{m^2}{B}\Big)^k-\\
    -\frac{m^4}{32\pi^2}\Big(\ln\frac{2m^2}{B}-\frac{3}{2}+\gamma\Big)-\frac{Bm^2}{16\pi^2}\ln2+\frac{B^2}{96\pi^2}\Big(\ln\frac{m^2}{4B}-12\zeta'(-1)\Big),
\end{multline}
which coincides exactly with formula (1.62) of \cite{Dunne2} for the effective Lagrangian.

\subsection{Dirac particles}

As for Dirac particles, we use the fermionic expansion, which in the case of strong fields and after the pole cancelation in the $\nu$ plane becomes
\begin{equation}
\begin{split}
    -\frac{\Omega_f(\mu)}{V}&=\sideset{}{'}\sum^{\infty}_{k,s,n=0}\Gamma(4-2k-2s)\eta(4-2k-2s-n)\beta^{-(4-2k-2s)} \frac{(\beta\mu)^n}{n!}\frac{(-1)^{k+s}m^{2s}a_k}{(4\pi)^{3/2}s!\Gamma(\frac{5}{2}-k-s)}+\\
    &+\sideset{}{'}\sum_{l,s,k=0}^{\infty}\frac{B^2}{(4\pi)^{3/2}}\frac{\eta(-l)2^{2+s-k} \Gamma(k-s-1)\zeta(k-s-1)}{(2s)!(l-2s)!k!\Gamma(\tfrac{1}{2}-s)}(\beta\mu)^l\Big(\frac{B}{\mu^2}\Big)^{s}\Big(-\frac{m^2}{B}\Big)^k+\\
    &+\sum_{k,l=0}^{\infty}\frac{(-1)^k\eta(-l)B\beta^l\mu^{l-2k+2}m^{2k}}{(4\pi)^{3/2}k!(2k-2)!(l-2k+2)!\Gamma(\frac{3}{2}-k)} \Big[\ln\frac{Be^\ga}{\pi m^2}+\psi(k+1)\Big]-\\
    &-\sideset{}{'}\sum^{\infty}_{n,s,k=0} \frac{(-1)^s\eta(-2s-n)}{16\pi^{3/2}(2-k+s)!\Gamma(\tfrac{1}{2}-s)} \frac{(\beta\mu)^n(\beta m)^{2s}}{n!(2s)!} m^{4-2k}a_k\times\\
    &\times\Big[\ln(2\be^2B)+\psi(\tfrac{1}{2}-s)-2\frac{\zeta'(-2s-n)}{\zeta(-2s-n)}-2\psi(2s+1)-\frac{\zeta'(1-k)}{\zeta(1-k)}-\psi(k) -\frac{4\ln 2}{2^{-n-2s}-2}\Big].
\end{split}
\end{equation}
Here the same notation and conventions are implied as in \eqref{omega_b}. The finite and divergent parts of the high-temperature expansion read as
\begin{multline}
    \frac{\Omega_f(\mu)}{V}=-\frac{7\pi^2}{360}\be^{-4}-\frac{3\zeta(3)}{2\pi^2}\mu\be^{-3}+\Big(\frac{m^2}{24}-\frac{\mu^2}{12}\Big)\be^{-2} +\frac{\ln2}{\pi^2}\Big(\frac{m^2}{2}-\frac{\mu^2}{3}\Big)\mu\be^{-1}+\frac{\mu^2}{8\pi^2}\Big(m^2-\frac{\mu^2}{3}\Big)-\\
    -\frac{B^2}{4\pi^2}\sum^{\infty}_{k=3}\Big(-\frac{m^2}{2B}\Big)^k\frac{\zeta(k-1)}{k(k-1)} +\Big(\frac{m^4}{32\pi^2}+\frac{B^2}{48\pi^2}\Big) \ln\frac{2\be^2Be^\ga}{\pi^2} +\frac{B^2}{48\pi^2}[12\zeta'(-1)+\ga-1]+\frac{m^2 B}{16\pi^2}\ln \frac{Be}{\pi m^2}.
\end{multline}
The contribution of charged Dirac fermions to the non-renormalized energy of vacuum fluctuations takes the form
\begin{multline}
    \frac{E_{vac}}{V}=-2\frac{\partial_{\beta}(\beta\Omega_f(0))}{V}=-\frac{7\pi^2}{60}\beta_0^{-4}+\frac{m^2}{12}\beta_0^{-2}-\Big(\frac{m^4}{8\pi^2}+\frac{B^2}{12\pi^2}\Big)\ln(e\beta_0)+\\
    +\frac{B^2}{2\pi^2}\sum^{\infty}_{k=3}\frac{\zeta(k-1)}{k(k-1)}\Big(-\frac{m^2}{2B}\Big)^k -\Big(\frac{m^4}{16\pi^2}+\frac{B^2}{24\pi^2}\Big) \ln\frac{2Be^\ga}{\pi^2} -\frac{B^2}{24\pi^2}[12\zeta'(-1)+\ga-1]-\frac{m^2 B}{8\pi^2}\ln \frac{Be}{\pi m^2}.
\end{multline}
In the case of weak fields, we have
\begin{equation}
\begin{split}
    -\frac{\Omega_f(\mu)}{V}&=\sideset{}{'}\sum_{k,s,n=0}^{\infty}\Gamma(4-2k-2s)\eta(4-2k-2s-n)\beta^{-(4-2k-2s)} \frac{(\beta\mu)^n}{n!}\frac{(-1)^{k+s}m^{2s}a_k}{(4\pi)^{3/2}s!\Gamma(\tfrac{5}{2}-k-s)}-\\
    &-\sum_{l,s=0}^{\infty}\sideset{}{'}\sum_{k=-1}^{k_{max}}\frac{B^2}{(4\pi)^{3/2}}\frac{\eta(-l)(\beta\mu)^l}{(2s)!(l-2s)!} \frac{\Gamma(2k-s)}{\Gamma(\tfrac{1}{2}-s)}\Big(\frac{B}{m^2}\Big)^{2k}\Big(\frac{m}{\mu}\Big)^{2s} 2^{2k+2}\frac{\zeta(-2k-1)}{\Gamma(2k+2)}-\\
    &-\sideset{}{'}\sum_{\substack{k,s,n=0\\2+s\geq k}}^{\infty} \frac{(-1)^s\eta(-2s-n)}{16\pi^{3/2}(2-k+s)!\Gamma(\tfrac{1}{2}-s)} \frac{(\beta\mu)^n(\beta m)^{2s}}{n!(2s)!} m^{4-2k}a_k\times\\
    &\times\Big[\ln(\beta^2m^2)+\psi(\tfrac{1}{2}-s)-2\frac{\zeta'(-2s-n)}{\zeta(-2s-n)} -\frac{4\ln 2}{2^{-n-2s}-2} -2\psi(2s+1)-\psi(3-k+s)\Big],
\end{split}
\end{equation}
without taking into account the non-perturbative corrections. The finite and divergent parts are written as
\begin{multline}
    \frac{\Omega_f(\mu)}{V}=-\frac{7\pi^2}{360}\be^{-4} -\frac{3\zeta(3)}{2\pi^2}\mu\be^{-3}+\Big(\frac{m^2}{24}-\frac{\mu^2}{12}\Big)\be^{-2} +\frac{\ln2}{\pi^2}\Big(\frac{m^2}{2}-\frac{\mu^2}{3}\Big)\mu\be^{-1} +\frac{\mu^2}{8\pi^2}\Big(m^2-\frac{\mu^2}{3}\Big)+\\
    +\Big(\frac{B}{2\pi}\Big)^2\sum_{k=1}^{k_{max}}\frac{\zeta(-2k-1)}{2k(2k+1)}\Big(\frac{2B}{m^2}\Big)^{2k} +\Big(\frac{m^4}{32\pi^2}+\frac{B^2}{48\pi^2}\Big) \ln\frac{\be^2m^2e^{2\ga}}{\pi^2} -\frac{3m^4}{64\pi^2}.
\end{multline}
In particular, the density of electrons at $\mu=0$ is given by (see, e.g., \cite{KhalilB}, Chap. 6)
\begin{equation}
    \frac{N}{V}=\frac{3\zeta(3)}{2\pi^2}\be^{-3},
\end{equation}
up to the terms vanishing at $\be\rightarrow0$. The non-renormalized vacuum contribution of charged Dirac fermions is
\begin{multline}
    \frac{E_{vac}}{V}=-2\frac{\partial_{\beta}(\beta\Omega_f(0))}{V}=-\frac{7\pi^2}{60}\beta_0^{-4}+\frac{m^2}{12}\beta_0^{-2}-\Big(\frac{m^4}{8\pi^2}+\frac{B^2}{12\pi^2}\Big)\ln\frac{\beta_0me^{\ga+1}}{\pi}+ \frac{3m^4}{32\pi^2}-\\
    -\frac{B^2}{2\pi^2}\sum^{k_{max}}_{k=1}\frac{\zeta(-2k-1)}{2k(2k+1)}\Big(\frac{2B}{m^2}\Big)^{2k} +\frac{(-1)^{[\pi m^2/2B]}}{4\pi^2}\Big(\frac{B}{\pi}\Big)^2 \Big(\frac{B}{2m^2}\Big)^{1/2}e^{-\pi m^2/B}.
\end{multline}
The counterterms to the initial Lagrangian are chosen by using the same rules as in \eqref{counterterms}:
\begin{equation}
    c.t.=-\frac{7\pi^2}{60}\beta_0^{-4}+\frac{m^2}{12}\beta_0^{-2}-\Big(\frac{m^4}{8\pi^2}+\frac{B^2}{12\pi^2}\Big)\ln\frac{\beta_0me^{\ga+1}}{\pi} + \frac{3m^4}{32\pi^2}.
\end{equation}
Therefore, in the weak field limit, the renormalized vacuum contribution takes the form
\begin{equation}
    \frac{E^{ren}_{vac}}{V}=\frac{E_{vac}}{V}-c.t.=-\frac{B^2}{2\pi^2}\sum^{k_{max}}_{k=1}\frac{\zeta(-2k-1)}{2k(2k+1)}\Big(\frac{2B}{m^2}\Big)^{2k}+\frac{(-1)^{[\pi m^2/2B]}}{4\pi^2}\Big(\frac{B}{\pi}\Big)^2 \Big(\frac{B}{2m^2}\Big)^{1/2}e^{-\pi m^2/B},
\end{equation}
which coincides exactly with formula (1.19) of \cite{Dunne2} for the effective Lagrangian without the non-perturbative contribution. As for the strong fields, we deduce
\begin{multline}
    \frac{E^{ren}_{vac}}{V}=\frac{E_{vac}}{V}-c.t.=\frac{B^2}{2\pi^2}\sum^{\infty}_{k=3}\frac{\zeta(k-1)}{k(k-1)}\Big(-\frac{m^2}{2B}\Big)^k+\\
    +\Big(\frac{m^4}{16\pi^2}+\frac{B^2}{24\pi^2}\Big)\ln\frac{m^2}{2B}+\frac{m^2 B}{8\pi^2}\ln\frac{\pi m^2}{B}+\frac{m^4}{16\pi^2}\Big(\gamma-\frac{3}{2}\Big)-\frac{m^2 B}{8\pi^2}+\frac{B^2}{24\pi^2}\Big(1-12\zeta'(-1)\Big),
\end{multline}
which coincides exactly with formula (1.53) of \cite{Dunne2} for the effective Lagrangian.

\section{Charged bosons in the magnetic field}\label{Char_Bos_Mag_Field}

Now we employ the high-temperature expansions obtained to analyze the main thermodynamic properties of a charged scalar boson gas in a constant homogeneous magnetic field at finite temperature and non-zero chemical potential. So, we assume
\begin{equation}\label{hte_region}
    \be m\ll1,\qquad \be^2 |B|\ll1,\qquad \be|\mu|\ll1.
\end{equation}
In this section, we neglect the contribution of photons to the omega-potential. Also we neglect the change of the effective masses of particles due to the contributions of the ring diagrams. These factors will be taken into account in the next section.

In the high-temperature limit, the leading contribution to the pressure can be cast into the form
\begin{equation}\label{pressure}
    P=-\Omega/V\approx\frac{\pi^2}{45}T^4+\Big(\frac{\mu^2}{6}-\frac{m^2}{12}\Big)T^2+TB^{3/2}\Phi\Big(\frac{\mu^2-m^2}{B}\Big),\qquad \Phi(x):=-\frac{\zeta\big(-1/2,(1-x)/2\big)}{\sqrt{2}\pi},
\end{equation}
where $T:=\be^{-1}$, the contributions of particles and antiparticles are taken into account, and the formulas \eqref{omega_bs}, \eqref{zeta_hurw} have been used. In the limit considered, the vacuum contribution can be neglected. The function $\Phi(x)$ is real for $x<1$ and possesses the square root branch point at $x=1$:
\begin{equation}
    \Phi(x)=\frac{\zeta(3/2)}{4\sqrt{2}\pi^2}-\frac{\sqrt{1-x}}{2\pi}+O(x-1).
\end{equation}
The condition $x<1$ is equivalent to $|\mu|<\omega_0:=\sqrt{m^2+B}$. The chemical potential is found from the equation
\begin{equation}\label{density}
    \rho=\frac{\partial P}{\partial\mu}\approx\frac{\mu T^2}{3}+2\mu TB^{1/2}\Phi'(x),
\end{equation}
where $x:=(\mu^2-m^2)/B<1$ and $\rho=Q/V$ is the charge density. The last term in \eqref{density} can dominate in the region \eqref{hte_region} provided $x\rightarrow1$. It is in this parameter domain where one ought to expect the phase transition. The higher terms of the high-temperature expansion discarded in \eqref{pressure}, \eqref{density} are regular in this limit and for other values of the chemical potential. Hence, they can be safely neglected.

The magnetic induction $B$ in the gas is determined by the equation
\begin{equation}\label{hb1}
    H=B-e^2\frac{\partial P}{\partial B}(\mu,T,B)=B-e^2TB^{1/2}[\tfrac32\Phi(x)-x\Phi'(x)],
\end{equation}
where $H$ is the magnetic intensity vector. Recall that the electric charge $e$ is included into the definition of the electromagnetic field strength, and we work in the system of units where $e^2=4\pi\al$ with $\al$ being the fine structure constant. Without loss of generality, we also assume $B>0$ and bear in mind that
\begin{equation}
    P(\mu,T,\mathbf{B})=P(\mu,T,-\mathbf{B}),
\end{equation}
in virtue of the time reversal symmetry. This relation is non-perturbative and valid in any order of the perturbation theory. In particular, $H(B)=-H(-B)$.

It is interesting to consider the behaviour of the chemical potential for the isochoric $\rho=const$ and adiabatic $s:=S/Q=const$ processes. In the first case, it approximately follows from \eqref{density} that
\begin{equation}
    \mu\approx3\rho\be^2,\;\;\rho\ll\omega_0T^2/3;\qquad\mu\approx\omega_0\Big[1-\frac{B^2T^2}{8\pi^2(\rho-\omega_0T^2/3)^2}\Big],\;\;\frac{B^2T^2}{8\pi^2(\rho-\omega_0T^2/3)^2}\ll1.
\end{equation}
In order to find the adiabatic curve, it is necessary to express $\mu=\mu(T,P,B)$ from \eqref{pressure}. Then the adiabatic equation becomes
\begin{equation}\label{entr_dens}
    s=-\frac{\partial\mu}{\partial T}(T,P,B)=const.
\end{equation}
Differentiating \eqref{pressure} with respect to $T$, we obtain the adiabatic equation
\begin{equation}\label{adiabatics}
    \frac{4\pi^2}{45}T^2-\frac{s\mu}{3}T-2s\mu B^{1/2}\Phi'(x)=0.
\end{equation}
Of course, one can solve exactly this equation with respect to $T$. However, we give here only the asymptotes for sufficiently small and large temperatures. If $x$ is not close to unity such that the last term in \eqref{adiabatics} can be neglected, then
\begin{equation}
    \mu\approx\frac{4\pi^2}{15 s}T.
\end{equation}
If $x$ is close to unity, then
\begin{equation}
    \mu^2\approx\omega_0^2\Big[1-\Big(\frac{45sB}{8\pi^3T^2}\Big)^2\Big],\qquad\Big(\frac{45sB}{8\pi^3T^2}\Big)^2\ll1.
\end{equation}
It is clear that $|\mu|<\omega_0$ as it should be. Both in the first and second cases
\begin{equation}
    V^{1/3}T\approx const,\qquad PV^{4/3}\approx const,
\end{equation}
on the adiabatic curve.

\begin{figure}[t]
\centering
\includegraphics*[width=0.4\linewidth]{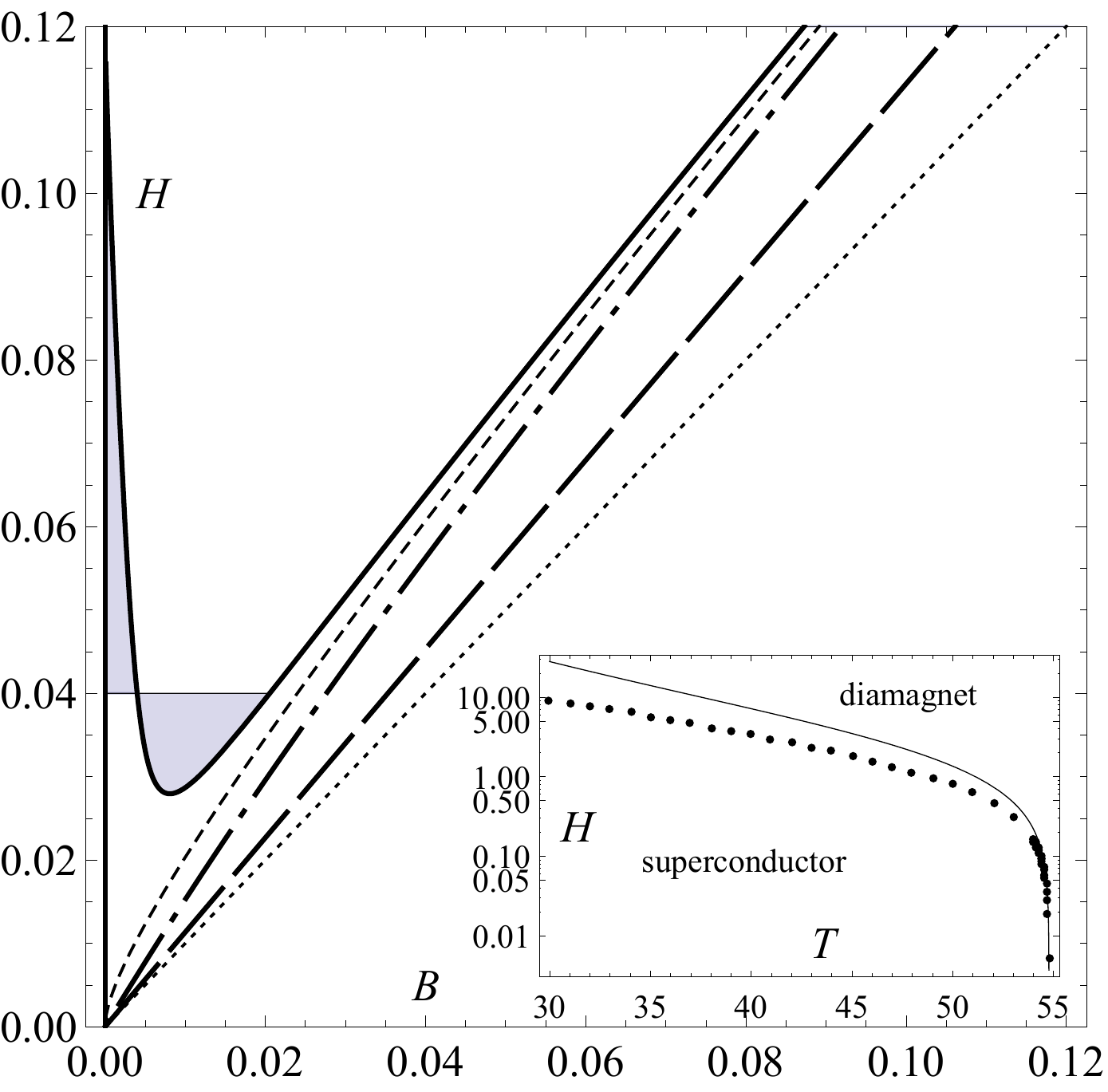}
\caption{{\footnotesize The dependence $H(B)$ for different temperatures at the charge density $\rho=1000$. The critical temperature at the vanishing magnetic field $T_c:=\sqrt{3\rho/m}\approx54.77$. The system of units is chosen such that $m=1$. The thick solid line is $H(B)$ for $T=54.7$. In this case, $H_1\approx0.12$ and $H_2\approx0.0028$. The thin solid line shows the Maxwell construction. The thin dashed line is $H(B)$ given in \eqref{hdia}. The thick dot-dashed line is $H(B)$ for $T=55$. The thick dashed line is $H(B)$ for $T=60$. The dotted line is $H=B$. The inset: The dependence $H_2(T)$ at $\rho=1000$. This dependence can be approximately considered as the equilibrium curve in the $(H,T)$ plane of the diamagnetic (above the curve) and superconducting (below the curve) phases. The solid line is the approximation presented in \eqref{H2_approx}. The dots denote the results of the numerical simulations.}}
\label{plotHB}
\end{figure}

The plot of $H(B)$ is given in Fig. \ref{plotHB} for the different temperatures. One can see from this plot that, at the sufficiently low temperature (but in the high-temperature limit \eqref{hte_region}), the system suffers the usual first-order phase transition, and the gas of charged bosons in the external magnetic field goes to the superconducting state. The dependence $H(B)$ shown in Fig. \ref{plotHB}. is typical for the superconductors of the first type (see \cite{LandLifECM}, Sect. 56). Thus we see that, at the non-vanishing \emph{external} magnetic field, $H\neq0$, the Bose-Einstein condensation of charged scalar bosons without self-interaction is possible \cite{Schafroth}. At that, the magnetic field is expulsed from the condensate so that \emph{locally}, i.e. in the condensate, $B=0$. This is the standard pattern of transition from the normal to the superconducting state (\cite{LandLifECM}, Sect. 57). So, our conclusion is in agreement with the conclusion made in \cite{ELPS}. We ought to add that the configuration of the intermediate state is determined by the minimum of $\mu(T,P,B)$ with the account of the Maxwell equations and is rather non-trivial (\cite{LandLifECM}, Sect. 57). The condensate does not fill homogeneously all the space, and its wave function is not the ground state of the Klein-Gordon equation in the homogeneous magnetic field.

Let us find the approximate explicit expressions for the main characteristics of this phase transition. The value $H_1:=H(0)$ (see Fig. \ref{plotHB}) can be readily found from \eqref{density}, \eqref{hb1} at $\rho>mT^2/3$ and $B\rightarrow0$. In this case, $\mu\rightarrow m$ and \cite{Schafroth,DFGK}
\begin{equation}
    H_1=\frac{e^2}{2m}\Big(\rho-\frac{mT^2}{3}\Big).
\end{equation}
The magnitude of the magnetic induction $B_0$ corresponding to the magnetic intensity $H_2$ (the value of $H$ at the extremum of the function $H(B)$) is found from the equation $H'(B_0)=0$, which is equivalent to
\begin{equation}\label{h2}
    \frac{\partial^2P}{\partial\mu^2}=e^2\Big[\frac{\partial^2P}{\partial\mu^2}\frac{\partial^2P}{\partial B^2}-\Big(\frac{\partial^2P}{\partial\mu\partial B}\Big)^2 \Big],\qquad P=P(\mu,T,B),
\end{equation}
where, having differentiated, one has to put $\mu=\mu(T,\rho,B)$ taken from \eqref{density}. Equation \eqref{h2} can be written as
\begin{equation}\label{phase_trn}
    \Phi(x)-\tfrac43 x\Phi'(x)+\tfrac43 x^2\Phi''(x)=\frac{4B_0^{1/2}}{3e^2T}+\frac{4\mu^2B_0^{-1/2}[\Phi'(x)-2x\Phi''(x)]^2}{T+6B_0^{1/2}[\Phi'(x)+2\mu^2 B_0^{-1}\Phi''(x)]}.
\end{equation}
Solving this equation with respect to $B_0$ and substituting its solution to \eqref{hb1}, we obtain the dependence $H_2(T,\rho)$. The numerical solution is presented in Fig. \ref{plotHB}. If the temperature is so high that
\begin{equation}
    \frac{4 B_0^{1/2}}{3e^2T}\ll1,
\end{equation}
then equation \eqref{phase_trn} is approximately reduced to
\begin{equation}
    \Phi(x)-\tfrac43 x\Phi'(x)+\tfrac43 x^2\Phi''(x)\approx0.
\end{equation}
The solution of this equation is $x_0\approx0.366$. Hence, $(3\rho/T^2)^2\approx\mu_0^2=m^2+x_0B_0$, and
\begin{equation}
    c_0:=\tfrac32\Phi(x_0)-x_0\Phi'(x_0)\approx-0.030.
\end{equation}
Then it follows from \eqref{hb1} that
\begin{equation}\label{H2_approx}
    H_2\approx \frac{m^2}{x_0}\Big[\Big(\frac{3\rho}{mT^2}\Big)^2-1\Big]-e^2mTc_0x_0^{-1/2}\Big[\Big(\frac{3\rho}{mT^2}\Big)^2-1\Big]^{1/2}.
\end{equation}
The comparison of this formula with the numerical solution is given in Fig. \ref{plotHB}. The curve $H_2(T,\rho)$ on the plane $(H,T)$ can be approximately regarded as the equilibrium curve of the diamagnetic and superconducting phases. Notice that one can reach the superconducting state moving along the adiabat \eqref{adiabatics} towards the increase of temperature (cf. \cite{HabWeld}).

For the sake of completeness, we present here the dependence $H(B)$ in the so-called super-diamagnetic regime \cite{DFGK,ELPS,DaicFr96}. One can easily deduce from \eqref{density}, \eqref{hb1} that \cite{DaicFr96}
\begin{equation}\label{hdia}
    H\approx B +3e^2 TB^{1/2}\frac{\zeta(3/2)}{(4\pi)^2}(\sqrt{2}-1),\quad\bigg|\frac{9\rho^2/T^4-m^2}{B}\bigg|\ll1.
\end{equation}
This formula describes quite well the dependence $H(B)$ for sufficiently large $B$, $B/T^2\ll1$.

\section{Ring diagrams}\label{Ring_Diag}

In the previous section, we have investigated in detail the one-loop omega-potential of the system of charged bosons. However, we did not take account of the fact that, at high temperatures, the effective masses of particles are changed considerably due to the infrared contributions of the diagrams of higher order in the coupling constant (see, e.g., \cite{KirzhLind,Linde,Kapustp,DolJack}). In order to take correctly these infrared contributions into account, one needs to sum an infinite number of the so-called ring diagrams \cite{DolJack}. As we shall see, the contributions of these diagrams change drastically the behaviour of the system at high densities and temperatures. Instead of the superconducting phase, the gas of charged bosons passes to the ferromagnetic state.

Let us consider the system of charged scalar particles with the self-interaction $\lambda \phi^4$ on a constant homogeneous magnetic field background. The Lagrangian density has the form
\begin{equation}\label{GL_model}
    \mathcal{L}=(\partial_{\mu}-i eA_{\mu})\Phi^{\ast}(\partial_{\mu}+i eA_{\mu})\Phi-m^2 |\Phi|^2-\lambda|\Phi|^4-\frac{1}{4}F_{\mu\nu}F^{\mu\nu},\quad\la>0,
\end{equation}
where $m^2>0$. To take into account the infrared contribution of the infinite number of the ring diagrams, it is convenient to seek for the effective masses of the fields self-consistently \cite{DolJack}. To this aim, we add and subtract the corresponding mass terms
\begin{equation}\label{lagr_dens}
    \mathcal{L}=(\partial_{\mu}-i eA_{\mu})\Phi^{\ast}(\partial_{\mu}+i eA_{\mu})\Phi-(m^2+m^2_{\chi}) |\Phi|^2-\lambda|\Phi|^4-\frac{1}{4}F_{\mu\nu}F^{\mu\nu}-\frac{1}{2}m^2_{\gamma}A^2_{0}+m^2_{\chi}|\Phi|^2+\frac{1}{2}m^2_{\gamma}A^2_{0}.
\end{equation}
The last two terms should be regarded as the interaction vertices and are taken into account as the perturbation. The quadratic part of the Lagrangian determining the propagators is defined with the account of the effective masses.

In order to find the pressure of the system considered, we represent the fields in the form
\begin{equation}
    \Phi(x)=\eta+\chi(x),\qquad A_{\mu}(x)=\bar{A}_{\mu}(x)+a_{\mu}(x),
\end{equation}
where $\eta=const\in\R$ characterizes the boson condensate. Henceforth, we consider the system in the parameter domain where $\eta=0$. However, we shall not set $\eta=0$ right away and find the pressure for the small constant $\eta$'s. This allows us to obtain the correction to the effective mass $m^2_\chi$ from the self-consistency equations \eqref{selfconsm}. As for $\bar{A}_{\mu}$, one should take $\bar{A}_{\mu}=(0,\mathbf{A})$, where $\mathbf{A}$ is the vector potential of the constant homogeneous magnetic field. Moreover, in quantum field theory at finite temperature and density, the chemical potential conjugate to the electromagnetic charge $Q$ enters into the Lagrangian density exactly as the zeroth component of the electromagnetic potential (see, e.g., \cite{Kapustp}). Therefore, it is convenient to put $\bar{A}_{\mu}=(\mu/e,\mathbf{A})$ and conduct all the calculations as for the zero chemical potential.

The corrections to masses can be found self-consistently as the second derivatives of the quantum correction to the effective action (in fact, to the pressure of the system $P$) with respect to the fields:
\begin{equation}\label{selfconsm}
    m^2_{\chi}=-\frac{1}{2}\frac{\partial^2 P(m_{\chi},m_{\gamma})}{\partial \eta^2}\Big|_{\eta=0,e\bar{A}_0=\mu}, \qquad m^2_{\gamma}=\frac{\partial^2 P(m_{\chi},m_{\gamma})}{\partial \bar{A}_{0}^2}\Big|_{\eta=0,e\bar{A}_0=\mu}.
\end{equation}
The following normalization conditions are assumed
\begin{equation}\label{norm_cond}
\begin{gathered}
    \frac{\partial m^2_\ga}{\partial \bar{A}_{0}}\Big|_{\eta=0,e\bar{A}_0=\mu}=\frac{\partial m^2_\chi}{\partial \bar{A}_{0}}\Big|_{\eta=0,e\bar{A}_0=\mu}=0,\\
    \frac{\partial^2 m^2_\ga}{\partial \bar{A}_{0}^2}\Big|_{\eta=0,e\bar{A}_0=\mu}=\frac{\partial^2 m^2_\chi}{\partial \bar{A}_{0}^2}\Big|_{\eta=0,e\bar{A}_0=\mu}=\frac{\partial^2 m^2_\ga}{\partial\eta^2}\Big|_{\eta=0,e\bar{A}_0=\mu}=\frac{\partial^2 m^2_\chi}{\partial\eta^2}\Big|_{\eta=0,e\bar{A}_0=\mu}=0.
\end{gathered}
\end{equation}
These conditions can always be satisfied in virtue of the renormalization ambiguity \cite{BogShir}. From the physical point of view, these normalization conditions say that the additional terms in \eqref{lagr_dens} renormalize only the masses of particles.

Let us find the propagators of the theory by isolating the quadratic part of the Lagrangian \eqref{lagr_dens} without the last two terms,
\begin{equation}
\begin{split}
    \mathcal{L}_{quad}&=(\partial_{\mu}-ie\bar{A}_{\mu})\chi^*(\partial_{\mu}+ie\bar{A}_{\mu})\chi-ie\eta a_{\mu}(\partial^{\mu}\chi-\partial^{\mu}\chi^*)+e^2\eta^2a^2 +2e^2\eta\bar{A}_{\mu}a^{\mu}(\chi+\chi^*)\\
    &-(m^2+m^2_{\chi}+2\lambda\eta^2)\chi\chi^*
    -\lambda\eta^2(\chi+\chi^*)^2+\frac{1}{2}a_{\mu}\Box a^{\mu}+\frac{1}{2}m^2_{\gamma}a^2_{0}+\frac12(\partial^\mu a_\mu)^2.
\end{split}
\end{equation}
We shall work in the Feynman gauge. In that case, the gauge condition and the Faddeev-Popov matrix become
\begin{equation}
    f=\partial^{\mu}a_{\mu}+ie\eta(\chi-\chi^*),\qquad \delta_{\varepsilon}f=\Box\varepsilon+e^2\eta(2\eta+\chi+\chi^*)\varepsilon.
\end{equation}
The ghost and the gauge-fixing Lagrangian densities are written as
\begin{equation}
    \mathcal{L}_{gh}=c[\Box+2e^2\eta^2+e^2\eta(\chi+\chi^*)]\bar{P},\qquad \mathcal{L}_{gf}=-\frac{1}{2}f^2=-\frac{1}{2}(\partial^{\mu}a_{\mu})^2-ie\eta\partial^{\mu}a_{\mu}(\chi-\chi^*)+\frac{1}{2}e^2\eta^2(\chi-\chi^*)^2.
\end{equation}

\begin{table}
  \centering
  \begin{tabular}{|c|c|c|}
    \hline
    Name & Mass squared & \# \\\hline
    Vector & $2e^2\eta^2$ & $1$ \\
    Vector & $2e^2\eta^2+m^2_\ga$ & $1$ \\
    Vector & $2e^2\eta^2$ & $2$ \\
    Ghost & $2e^2\eta^2$ & $-2$ \\
    Scalar & $m^2+m^2_\chi+2\la\eta^2+e^2\eta^2$ & $1$ \\
    Scalar & $m^2+m^2_\chi+6\la\eta^2$ & $1$ \\
    \hline
  \end{tabular}
  \caption{{\footnotesize The spectrum of real and fictitious particles of the model \eqref{lagr_dens} in the Feynman gauge.}}\label{spectrum}
\end{table}

If the contribution of the vertex
\begin{equation}\label{mixing}
    2e^2\eta\bar{A}_{\mu}a^{\mu}(\chi+\chi^*)
\end{equation}
is negligible, then the photon sector completely decouples from the scalar one \cite{Kapustp}. The one-loop correction to the pressure is given by the ``thermal'' determinant. In the sector of the $\chi$ fields, it takes the form
\begin{equation}
    \det D_{\chi}^{-1}=
    \begin{vmatrix}
    \frac{\delta^2 S}{\delta\chi^2} & \frac{\delta^2 S}{\delta\chi \delta\chi^*}\\
    \frac{\delta^2 S}{\delta\chi \delta\chi^*} & \frac{\delta^2 S}{\delta\chi^{*2}}\\
    \end{vmatrix}=\det\Big(\frac{\delta^2 S}{\delta\chi^2}-\frac{\delta^2 S}{\delta\chi \delta\chi^*}\Big) \det\Big(\frac{\delta^2 S}{\delta\chi^2}+\frac{\delta^2 S}{\delta\chi \delta\chi^*}\Big).
    \end{equation}
The last equality holds since $\chi$ and $\chi^*$ enter symmetrically into $S$. Explicitly, we obtain
\begin{equation}\label{determ_1}
    \det D_{\chi}^{-1} =\det\big[(\partial^{\mu}+ie\bar{A}_{\mu})^2+m^2+m_{\chi}^2+2\lambda\eta^2+e^2\eta^2\big] \det\big[(\partial^{\mu}+ie\bar{A}_{\mu})^2+m^2+m_{\chi}^2+6\lambda\eta^2\big].
\end{equation}
Now it is easy to find the spectrum of particles in the model (see table \ref{spectrum}). This information is sufficient to find the one-loop correction to the pressure with the leading contribution from the ring diagrams at high temperatures. In the one-loop approximation, the last two terms in \eqref{lagr_dens} are taken into account only at the tree level.

Taking into account the mixing term \eqref{mixing}, the functional determinant \eqref{determ_1} is multiplied by
\begin{equation}\label{mixing_contr}
    \det(1-D_a V^T D_\chi V)\approx1- \Tr(D_a V^T D_\chi V),
\end{equation}
where $V$ denotes the second variational derivative of \eqref{mixing}, and $D_a$ is the photon propagator. Inasmuch as we put $\eta=0$ in the final answer, the term \eqref{mixing_contr} is important only in calculating the temperature correction \eqref{selfconsm} to the mass squared $m^2_\chi$. In the high-temperature limit, the contribution of the mixing \eqref{mixing_contr} is suppressed in comparison with the ``direct'' contributions of the particles presented in table \ref{spectrum}. This contribution contains the two propagators, at least, and the masses of the particles entering these propagators are proportional to $T^2$ in the high-temperature limit (see \eqref{mass_temp}).

According to table \ref{spectrum}, in the high-temperature limit, the leading contribution to the pressure comes from the two massive photon degrees of freedom
\begin{equation}
    P_1=\frac{\pi^2}{90}T^4-\frac{2e^2\eta^2}{24}T^2, \qquad P_2=\frac{\pi^2}{90}T^4-\frac{2e^2\eta^2+m_{\gamma}^2}{24}T^2,
\end{equation}
and from the charged scalar fields
\begin{equation}
\begin{split}
    P_3&=\frac{\pi^2}{90}T^4+\Big(\frac{\mu^2}{12} -\frac{m^2+m_{\chi}^2+2\lambda\eta^2+e^2\eta^2}{24}\Big)T^2+\frac{TB^{3/2}}{2}\Phi\Big(\frac{\mu^2-m^2-m_{\chi}^2-2\lambda\eta^2-e^2\eta^2}{B}\Big),\\
    P_4&=\frac{\pi^2}{90}T^4+\Big(\frac{\mu^2}{12} -\frac{m^2+m_{\chi}^2+6\lambda\eta^2}{24}\Big)T^2+\frac{TB^{3/2}}{2}\Phi\Big(\frac{\mu^2-m^2-m_{\chi}^2-6\lambda\eta^2}{B}\Big),
\end{split}
\end{equation}
where we include again the electric charge $e$ to the definition of the electromagnetic field strength. The ghosts cancel the contribution of the two additional massive photon degrees of freedom.

Making use of equations \eqref{selfconsm} with the account of the normalization conditions \eqref{norm_cond}, we obtain the equations for masses
\begin{equation}\label{mass_temp}
    m^2_{\gamma}=e^2\frac{T^2}{3}+2e^2TB^{1/2}\big[\Phi'(x)+\frac{2\mu^2}{B}\Phi''(x)\big],\qquad
    m^2_{\chi}=(8\lambda+5e^2)\frac{T^2}{24}+(8\lambda+e^2)\frac{TB^{1/2}}{2}\Phi'(x),
\end{equation}
where $x:=(\mu^2-m^2-m_{\chi}^2)/B$. Solving these equations, one can find the masses as functions of $T$, $B$, and $\mu$. Of course, one should keep in mind that formulas \eqref{mass_temp} are valid only in the high-temperature limit \eqref{hte_region}, where $m$ is the effective mass of the particle.

\begin{figure}[t]
\centering
\includegraphics*[width=0.45\linewidth]{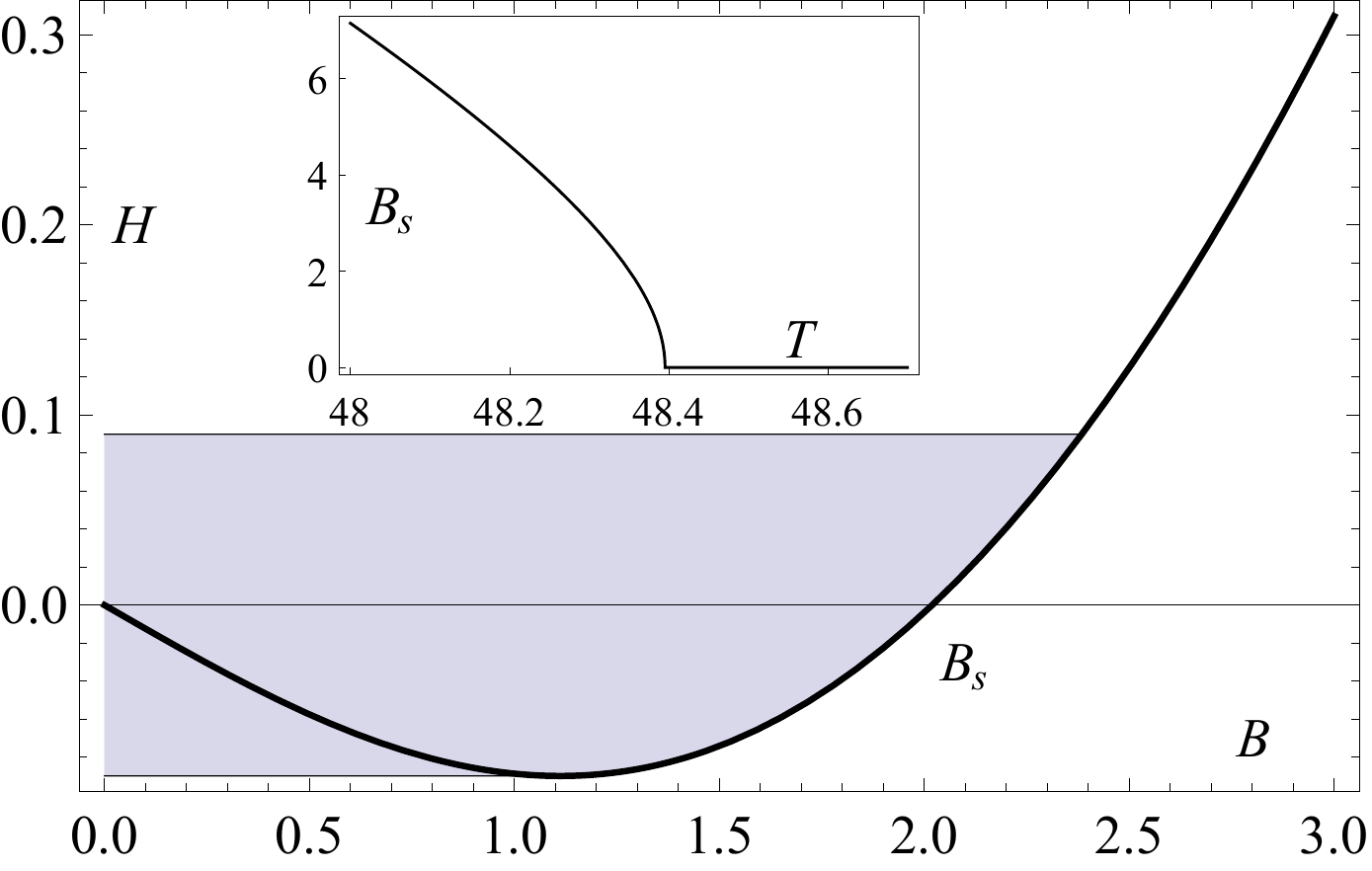}\;\;
\includegraphics*[width=0.45\linewidth]{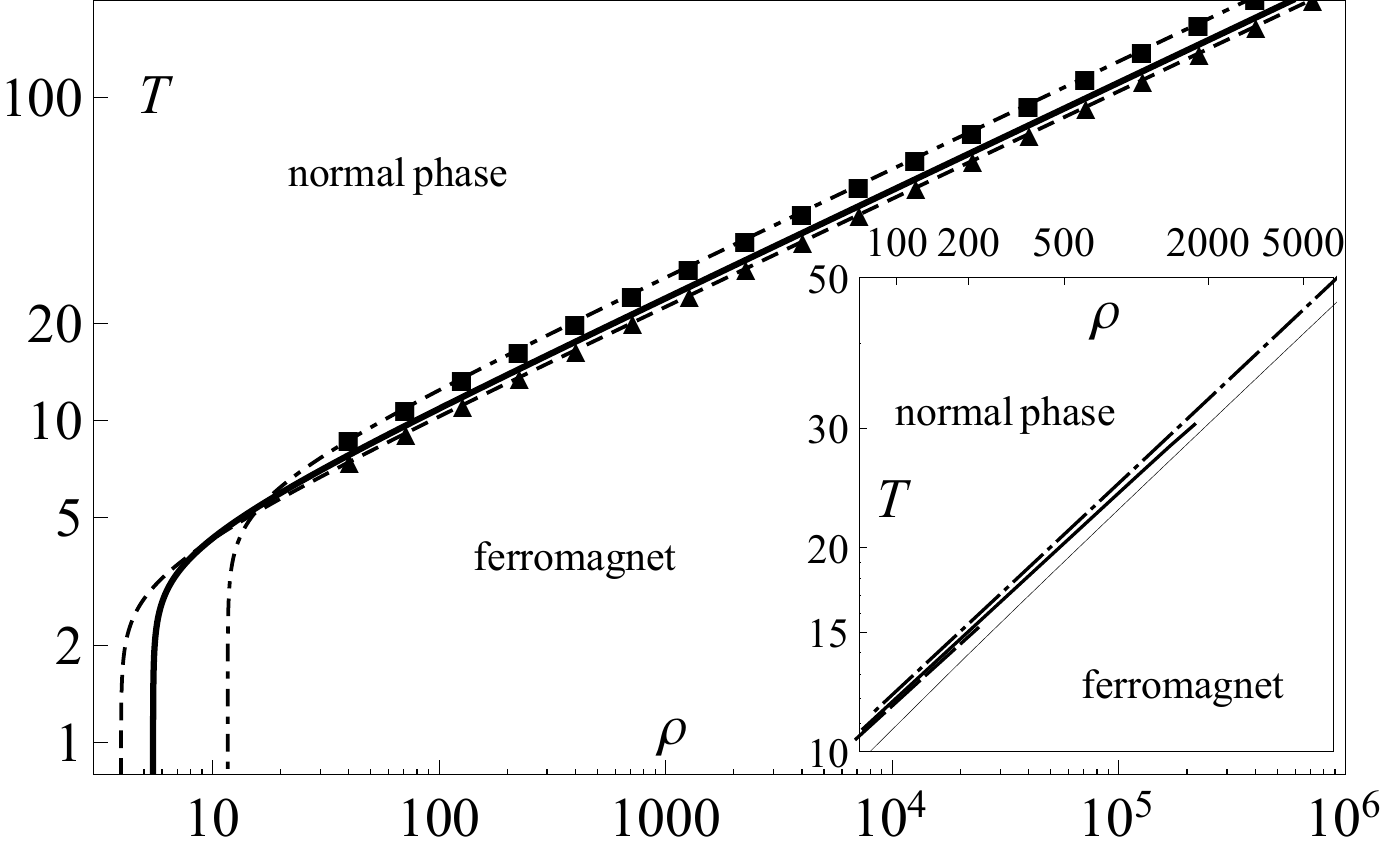}
\caption{{\footnotesize Left panel: The dependence $H(B)$ for $\la=1/10$, $\rho=8000$, $T=48.35$, and the Curie temperature $T_c=48.4$. The system of units is chosen such that $m=1$. The thick solid line is the theoretical dependence $H(B)$. It intersects the $x$ axis at the value of the spontaneous magnetization of the system. The filled region bounded by the two thin solid lines is a metastable region (the hysteresis loop), where the ferromagnetic domains can form in the gas. The inset: The dependence of the spontaneous magnetization on the temperature at $\rho=8000$. The numerical simulations reveal that the characteristic features of these plots (the order of the phase transition, the ferromagnetic state, etc.) do not depend on the value of the self-interaction coupling constant $0<\la\lesssim1/10$. Right panel: The dependence of the Curie temperature on the charge density $T_c(\rho)$. This curve can be regarded as the equilibrium curve of the ferromagnetic (below the curve) and the normal (above the curve) phases in the $(T,\rho)$ plane. The solid, dashed, and dot-dashed lines are $T_c(\rho)$ for $\la=1/10$, $\la=1/6$, and $\la=1/100$, respectively, given by formula \eqref{Curie_temp}. The marks on the curves are the results of the numerical simulations. Of course, these curves make sense only in the region $T\gtrsim10$. Inset: The adiabats $T(\rho,B)$ at $\la=1/10$ and $B=1$ are depicted. The thick dashed, solid, and dot-dashed lines are the adiabats for the entropies per unit charge $s=33$, $s=34$, and $s=36$, respectively. The thin solid line is the plot of the Curie temperature (the equilibrium curve). The thick dashed and solid adiabats terminate near the equilibrium curve, while the dot-dashed curve does not. This shows that the ferromagnetic state can be reached adiabatically provided $s$ is not too large.}}
\label{gist}
\end{figure}

For $\eta=0$, the total pressure can be cast into the form
\begin{equation}
    P=\frac{2\pi^2}{45}T^4+\frac{T^2}{12}\Big(2\mu^2-m^2-m_{\chi}^2-\frac{m_{\gamma}^2}{2}\Big)+T B^{3/2}\Phi(x).
\end{equation}
The chemical potential $\mu=\mu(T,\rho,B)$ is obtained from the definition of the charge density
\begin{equation}\label{charge_dens}
    \rho=\frac{\partial P}{\partial(e \bar{A}_0)}\Big|_{\eta=0,e\bar{A}_0=\mu}=\frac{T^2}{3}\mu+2TB^{1/2}\mu \Phi'(x).
\end{equation}
At last, the magnetization is written as
\begin{equation}\label{magnetiz}
    \frac{M}{e^2}=\frac{\partial P}{\partial B}=-\frac{T^2}{12}(\dot{m}_{\chi}^2+\tfrac{1}{2}\dot{m}_{\gamma}^2)+TB^{1/2}[\tfrac32\Phi(x)-(x+\dot{m}_{\chi}^2)\Phi'(x)],
\end{equation}
where the dot denotes the derivative with respect to $B$. Sequentially substituting into this expression $m^2_{\chi}=m^2_{\chi}(T, B, \mu)$ and $m^2_{\gamma}=m^2_{\gamma}(T, B, \mu)$, and then $\mu=\mu(T,\rho,B)$, we get the magnetization as the function of the variables $T$, $\rho$, and $B$. The magnetic field intensity is related to the induction and magnetization by the standard formula \eqref{hb1}. The plot of $H(B)$ is given in Fig. \ref{gist}.


The numerical study shows that the main contribution to the magnetization stems from the first term in \eqref{magnetiz} proportional to $T^2$, and
\begin{equation}\label{mdt}
    \dot{m}^2_\chi<0,\qquad \dot{m}^2_\ga<0.
\end{equation}
The effective magnetic moment of a particle can be defined as (\cite{LandLifstat}, Sect. 71)
\begin{equation}
    -\frac{\partial\e}{\partial B},
\end{equation}
where $\e$ is particle's dispersion law. In our case,
\begin{equation}\label{magn_mom}
    -\frac{\partial\e_\chi}{\partial B}=-\frac{2n+1+\dot{m}^2_\chi}{2\e_\chi},\qquad -\frac{\partial\e_\ga}{\partial B}=-\frac{\dot{m}^2_\ga}{2\e_\ga},
\end{equation}
where $n$ is the Landau level number. One can see from \eqref{mdt}, \eqref{magn_mom} that it is advantageous for the particles striving for a minimum energy to increase the magnetic field. Therefore, at sufficiently high density of charged bosons described by the model \eqref{GL_model}, the system has to pass to the ferromagnetic state. The numerical simulations confirm this observation. Formulas \eqref{mass_temp}, \eqref{charge_dens}, and \eqref{magnetiz} allow one to deduce a rough estimate for the Curie temperature
\begin{equation}\label{Curie_temp}
    T_c=\Big(\frac{24}{8\la+5e^2}\Big)^{1/6}\Big\{9\rho^2\Big[1+\Big(\frac{e^4}{64\pi}\Big)^{2/5}\Big(\frac{24}{8\la+5e^2}\Big)^{3/5}\Big]-m^2(3\rho)^{4/3}\Big(\frac{24}{8\la+5e^2}\Big)^{2/3} \Big\}^{1/6},
\end{equation}
which is in a rather well agreement with the numerical calculations. The first-order phase transition to the ferromagnetic state can happen only if the charge density
\begin{equation}
    \rho>\rho_c:=\frac{8m^3}{8\la+5e^2}\Big[1+\Big(\frac{e^4}{64\pi}\Big)^{2/5}\Big(\frac{24}{8\la+5e^2}\Big)^{3/5}\Big]^{1/3}.
\end{equation}
The numerical simulations show that, as in the previous section, where the ring diagrams were not taken into account, one can reach the phase transition domain moving along the adiabatic curve from the region of low temperatures provided the entropy per unit charge \eqref{entr_dens} is sufficiently small (see Fig. \ref{gist}). The form of $\eta(\spx)$ in the ferromagnetic state has the standard domain structure for ferromagnets as it is described, for example, in \cite{LandLifstat}, Sect. 44.


\section{Conclusion}

Thus, we derived the explicit formulas for the high-temperature expansion of the one-loop corrections to the omega-potential induced by both scalar and Dirac charged particles in a constant homogeneous magnetic field. These formulas generalize the known ones \cite{DFGK,CanDunne} in that respect that the contributions of particles and antiparticles are treated separately, i.e., one can find from them the number of particles and not only the total charge. Then we employed these formulas to describe the thermodynamic properties of a gas of charged bosons in a magnetic field at high temperatures and non-zero charge density. The two models were considered: with and without the contribution of the ring diagrams. The latter model was investigated in many papers \cite{Schafroth,DFGK,ELPS,PerRoj1,PerRoj2,PerRojVilLel,DelBarSol,StandTom2,Toms1995,DaicFr96,StandTom1}, and our conclusions are mainly agree with those given in \cite{ELPS}. In addition, we established that the system suffers the usual first-order phase transition from the normal to the superconducting state and found the equilibrium curve of these two phases in the magnetic field. As for the first model, we found that the contributions of photons and of the ring diagrams change drastically the behaviour of the system at high densities. Instead of the superconducting phase, the system passes into the ferromagnetic state. The main thermodynamic properties of this system were analyzed and the approximate formula for the Curie temperature was obtained.

\paragraph{Acknowledgments.}

The work of ISK is partly supported by the Ministry of Science of the Russian Federation (Grant No 2014/223), Code project 1766. The work of POK is supported by the RFBR grant No 16-02-00284.

\end{document}